\documentclass[longauth]{aa}
\usepackage{natbib}
\bibpunct{(}{)}{;}{a}{}{,} 
\usepackage{graphicx}
\usepackage{txfonts}
\usepackage{color}
\newcommand{\degree}{$^{\circ}$}

\usepackage[backref=true,pagebackref=true,hyperindex=true,breaklinks=true,
	colorlinks=true,urlcolor=blue,linkcolor=blue,citecolor=blue,
	bookmarks=true,bookmarksopen=true]{hyperref}

\begin{document} 

\title{New constraints on the disk characteristics and companion candidates around T~Cha with VLT/SPHERE\thanks{Based on observations made with European Southern Observatory (ESO) telescopes at the Paranal Observatory in Chile, under program IDs 095.C-0298(B), 096.C-0248(B) and 096.C-0248(C).}}

\author{A.~Pohl\inst{1,2} \and E.~Sissa\inst{3,4} \and M.~Langlois\inst{5,6} \and A.~M\"{u}ller\inst{1,7} \and C.~Ginski\inst{8,9}  \and R.~G.~van Holstein\inst{8} \and A.~Vigan\inst{6} \and D.~Mesa\inst{3} \and A.-L.~Maire\inst{1} \and Th.~Henning\inst{1} \and R.~Gratton\inst{3} \and J.~Olofsson\inst{10,1,11} \and R.~van~Boekel\inst{1} \and M.~Benisty\inst{12} \and B.~Biller\inst{1,13} \and A.~Boccaletti\inst{14} \and G.~Chauvin\inst{12} \and S.~Daemgen\inst{15} \and J.~de~Boer\inst{8} \and S.~Desidera\inst{3} \and C.~Dominik\inst{9} \and A.~Garufi\inst{15} \and M.~Janson\inst{1,16} \and Q.~Kral\inst{17} \and F.~M\'{e}nard\inst{12} \and C.~Pinte\inst{12} \and T.~Stolker\inst{9} \and J.~Szul\'{a}gyi\inst{15} \and A.~Zurlo\inst{6,18,19} \and M.~Bonnefoy\inst{12} \and A.~Cheetham\inst{20} \and M.~Cudel\inst{12} \and M.~Feldt\inst{1} \and M.~Kasper\inst{21} \and A.-M.~Lagrange\inst{12} \and C.~Perrot\inst{14} \and F.~Wildi\inst{20}
}

\institute{Max Planck Institute for Astronomy, K\"onigstuhl 17, D-69117 Heidelberg, Germany\\
	\email{pohl@mpia.de}
    \and
	Heidelberg University, Institute of Theoretical Astrophysics, Albert-Ueberle-Str. 2, D-69120 Heidelberg, Germany
	\and
	INAF-Osservatorio Astronomico di Padova, Vicolo dell’Osservatorio 5, 35122 Padova, Italy
	\and
	Dipartimento di Fisica e Astronomia "G. Galilei", Universita' degli Studi di Padova, Vicolo dell'Osservatorio 3, 35122 Padova, Italy
	\and
	CRAL, UMR 5574, CNRS, Universit\'{e} Lyon 1, 9 avenue Charles Andr\'{e}, 69561 Saint Genis Laval Cedex, France
	\and	
	Aix Marseille Universit\'{e}, CNRS, LAM (Laboratoire d'Astrophysique de Marseille) UMR 7326, 13388, Marseille, France
	\and
	European Southern Observatory, Alonso de C\'{o}rdova 3107, Casilla 19001 Vitacura, Santiago 19, Chile	
	\and
	Leiden Observatory, Leiden University, P.O. Box 9513, 2300 RA Leiden, The Netherlands
	\and
	Anton Pannekoek Institute for Astronomy, University of Amsterdam, Science Park 904, 1098 XH Amsterdam, The Netherlands	
	\and
	Instituto de F\'{i}sica y Astronom\'{i}a, Facultad de Ciencias, Universidad de Valpara\'{i}so, Av. Gran Breta\~{n}a 1111, Playa Ancha, Valpara\'{i}so, Chile
	\and
	ICM nucleus on protoplanetary disks, Protoplanetary discs in ALMA Early Science, Universidad de Valpara\'{i}so, Valpara\'{i}so, Chile	
	\and
	Univ. Grenoble Alpes, CNRS, IPAG, F-38000 Grenoble, France
	\and
	Institute for Astronomy, The University of Edinburgh, Royal Observatory, Blackford Hill View, Edinburgh, EH9 3HJ, UK
	\and
	LESIA, Observatoire de Paris, PSL Research University, CNRS, Sorbonne Universités, UPMC Univ. Paris 06, Univ. Paris Diderot, Sorbonne Paris Cité, 5 place Jules Janssen, 92195 Meudon, France
	\and
	Institute for Astronomy, ETH Zurich, Wolfgang-Pauli-Strasse 27, 8093 Zurich, Switzerland
	\and
	Department of Astronomy, Stockholm University, AlbaNova University Center, 10691 Stockholm, Sweden	
	\and
	Institute of Astronomy, University of Cambridge, Madingley Road, Cambridge CB3 0HA, UK
	\and
	N\'{u}cleo de Astronom\'{i}a, Facultad de Ingenier\'{i}a, Universidad Diego Portales, Av. Ejercito 441, Santiago, Chile
	\and
	Departamento de Astronom\'{i}a, Universidad de Chile, Casilla 36-D, Santiago, Chile
	\and
	Observatoire Astronomique de l’Universit\'{e} de Gen\`{e}ve, 51 Ch. des Maillettes, 1290 Versoix, Switzerland
	\and
	European Southern Observatory, Karl-Schwarzschild-Str. 2, 85748 Garching, Germany
	}


\abstract
   {The transition disk around the T Tauri star T~Cha possesses a large gap, making it a prime target for high-resolution imaging in the context of planet formation.}
   {We aim to find signs of disk evolutionary processes by studying the disk geometry and the dust grain properties at its surface, and to search for companion candidates.}
   {We analyze a set of VLT/SPHERE data at near-infrared and optical wavelengths. We performed polarimetric imaging of T~Cha with IRDIS (1.6\,$\mu$m) and ZIMPOL (0.5--0.9\,$\mu$m), and obtained intensity images from IRDIS dual-band imaging with simultaneous spectro-imaging with IFS (0.9--1.3\,$\mu$m).}
   {The disk around T~Cha is detected in all observing modes and its outer disk is resolved in scattered light with unprecedented angular resolution and signal-to-noise. The images reveal a highly inclined disk with a noticeable east-west brightness asymmetry. The significant amount of non-azimuthal polarization signal in the $U_{\phi}$ images, with a $U_{\phi}/Q_{\phi}$ peak-to-peak value of 14\%, is in accordance with theoretical studies on multiple scattering in an inclined disk. Our optimal axisymmetric radiative transfer model considers two coplanar inner and outer disks, separated by a gap of 0\farcs28 ($\sim$30\,au) in size, which is larger than previously thought. We derive a disk inclination of $\sim$69\,deg and PA of $\sim$114\,deg. In order to self-consistently reproduce the intensity and polarimetric images, the dust grains, responsible for the scattered light, need to be dominated by sizes of around ten microns. A point source is detected at an angular distance of 3.5\arcsec from the central star. It is, however, found not to be co-moving.}
  {We confirm that the dominant source of emission is forward scattered light from the near edge of the outer disk. Our point source analysis rules out the presence of a companion with mass larger than $\sim$8.5\,$M_{\mathrm{jup}}$ between 0\farcs1 and 0\farcs3. The detection limit decreases to $\sim$2\,$M_{\mathrm{jup}}$ for 0\farcs3 to 4.0\arcsec.}
  
%
%
\titlerunning{VLT/SPHERE observations of the disk around T~Cha}
\maketitle

%
%
\section{Introduction}
\label{sec:introduction}
Recently developed high-resolution and high-contrast imaging instruments provide the excellent capability to directly obtain images of protoplanetary disks in scattered light and thermal emission. Protoplanetary disks are optically thick in the optical and near-infrared (NIR), so that scattered light imaging probes (sub-)micron-sized dust grains in the disk surface layer. Contrarily, (sub-)mm observations trace larger, mm-sized grains located in the disk midplane. The detection of (non-)axisymmetric disk features is fundamental in improving our current understanding of disk evolution and planet formation. Transition disks with large gas and dust gaps (see e.g. mid-infrared surveys by \citealt{brown2007,merin2010,vandermarel2016} and mm observations by \citealt{isella2010a}, \citeyear{isella2010b}; \citealt{andrews2011,vandermarel2015}) are particularly interesting targets, since they host possible planet-forming hotspots and may show signposts of planet-disk interaction processes. In recent observational studies, giant gaps and cavities have been directly imaged in scattered light observations of transition disk systems (e.g., \citealt{thalmann2010}, \citeyear{thalmann2015}; \citealt{hashimoto2012,avenhaus2014,follette2015,ohta2016,stolker2016,benisty2017}).\\

T~Chamaeleontis (T~Cha) is a $\sim$2--12\,Myr old (\citealt{brown2007,torres2008,murphy2013}) T Tauri star (spectral type G8, \citealt{alcala1993}) at an estimated distance of $107 \pm 3\,$pc (first GAIA data release, \citealt{gaia2016a}) surrounded by a transition disk. Several recent studies covering a wide wavelength range dealt with constraining the disk geometry around T~Cha. \citet{brown2007} studied its spectral energy distribution (SED), which features a significant deficit of mid-infrared (MIR) emission. With radiative transfer based SED fitting they modeled this deficit by introducing a gap between 0.2 and 15\,au, dividing the disk into two spatially separated parts. Hence, it became a prime candidate for investigating signatures of ongoing planet formation. \citet{olofsson2011} presented spatially resolved, interferometric observations at high angular resolution in the NIR from the AMBER instrument at the Very Large Telescope Interferometer (VLTI) to study the inner disk's structure. The inner disk is found to be extremely narrow and located close to the star with an extension from 0.13 to 0.17\,au. \citet{olofsson2013} presented a radiative transfer model accounting for several further interferometric and photometric observations, including VLTI/PIONIER, VLTI/MIDI and NACO/Sparse Aperture Masking (SAM) data, which further constrains the inner disk to extend from 0.07 to 0.11\,au. Further SED modeling of T~Cha by \citet{cieza2011} suggests that there is a high degeneracy especially for the outer disk geometry, since a very compact outer disk provides an equally good fit to the Herschel data as a much larger, but tenuous disk with a very steep surface density profile. High-resolution Atacama Large Millimeter/sub-millimeter Array (ALMA) observations of the 850\,$\mu$m dust continuum as well as of several emission lines presented by \citet{huelamo2015} spatially resolved the outer disk around T~Cha and helped to break the degeneracy of previous outer disk models. They report a compact dusty disk, where the continuum intensity profile displays two emission bumps separated by 40\,au, indicating an inner gap size of 20\,au and an outer disk radius of $\sim$80\,au. In contrast, the gaseous disk is larger by almost a factor of three, giving a radius of $\sim$230\,au based on the detection of CO(3--2) molecular emission. \citet{huelamo2015} derived a disk inclination (incl) of $67^{\circ} \pm 5^{\circ}$ and a position angle (PA) of $113^{\circ} \pm 6^{\circ}$ by fitting a Gaussian to the CO(3--2) integrated emission map.\\

All previous observations clearly confirm that there must be a significant gap in the disk dust density distribution, while its origin is still debated. In general, radial gap structures can be created by a number of processes, including grain growth (e.g., \citealt{dullemond2005}), effects of the magneto-rotational instability (MRI) at the outer edge of a dead-zone (e.g., \citealt{flock2015}), a close (sub-)stellar companion or the dynamical interaction of a planet formed within the disk (e.g., \citealt{rice2003}). For the latter, the disk density modification results from the torques exerted on the disk by the planet and by the disk itself. The planet pushes away the surrounding material, the outer part of the disk outward and the inner part inward, thereby opening a gap (e.g., \citealt{lin1979,crida2006}). Studies of observational signatures of planet-disk interaction processes based on numerical simulations suggest that gaps detected in scattered light may be opened by planets (e.g., \citealt{pinilla2015,dong2015}, \citeyear{dong2016}; \citealt{juhasz2015,pohl2015}). Using NACO/SAM \citet{huelamo2011} detected a companion candidate at a projected distance of 6.7\,au from the primary, which is well within the previously described disk gap. However, an analysis of several L' and $K_{\mathrm{s}}$ data sets covering a period of three years ruled out this companion hypothesis (\citealt{olofsson2013,cheetham2015}). The absence of relative motion for the companion candidate favors a stationary structure consistent with scattered light from a highly inclined disk. \citet{sallum2015} checked if the closure phase signal from their VLT/NACO and Magellan/MagAO/Clio2 data shows any variation in time, which is not expected for the disk scattering model. While NACO L' data from 2011 and 2013 support the hypothesis of constant scattered light from the disk, the best fits for two other NACO data sets are inconsistent, requiring temporal variability in the amount of scattered light. Apart from this variability argument, \citet{sallum2015} showed by means of Monte Carlo simulations that noise fluctuations could also cause the changing structure in the NACO and MagAO reconstructed images.\\

In this work we present the first scattered light observations of T~Cha obtained with the SPHERE instrument (Spectro-Polarimeter High
contrast Exoplanet REsearch, \citealt{beuzit2008}) at the Very Large Telescope (VLT). The target is now one of the few T Tauri stars to have been spatially resolved in high detail in scattered light. We used Polarimetric Differential Imaging (PDI) complemented with total intensity images obtained with the angular differential imaging (ADI) technique. The observations provide the first spatially resolved high-contrast images of T~Cha in the optical and NIR. Our focus is set on analyzing the scattered light properties of the disk. We perform physical modeling of the disk via radiative transfer calculations, which helps us to further constrain the disk's geometry and grain size distribution. Images in the full Stokes vector are calculated in order to consistently reproduce the observed total and polarized intensity. Furthermore, the total intensity images are used for a detailed search for substellar companion candidates and, in case of non-detection, to place constraints on the mass of putative companions using the detection limits. This paper is laid out as follows. In Sect. \ref{sec:obs} we describe our observations and the data reduction procedures; their results are shown in Sect. \ref{sec:results}. Section \ref{sec:ana} presents our results of the radiative transfer disk modeling and the search for substellar companions. A detailed discussion in Sect. \ref{sec:discussion} follows. In Sect. \ref{sec:conclusions} we summarize the main conclusions of this work.

\section{SPHERE observations and data reduction}
\label{sec:obs}

\renewcommand{\arraystretch}{1.2}
\begin{table*}
	\caption{Overview of observational data sets}              
	\label{tab:obs}      
	\centering
	\begin{tabular}{cccccccccc}          
		\hline\hline                       
		Date & Instrument & Mode & Filter & DIT [s] $\times$ NDIT & PC & t$_{\mathrm{tot}}$ [min] & Seeing [$''$] & \textit{H}-band Strehl [\%]\\
		\hline
		2015 May 30 & IRDIS & IRDIFS & \textit{DB\_H2H3} & 64 $\times$ 96 & -- & 102 & 0.5--0.85 & 27$\pm$13\\
		2015 May 30 & IFS & IRDIFS & \textit{YJ} & 64 $\times$ 96 & -- & 102 & 0.5--0.85 & 27$\pm$13\\
		2016 February 19 & IRDIS & DPI & \textit{H} & 32 $\times$ 1 & 30 & 64 & 0.9--1.0 & 76$\pm$4\\
		2016 March 31 & ZIMPOL & P2 & \textit{VBB} & 40 $\times$ 2 & 18 & 96 & 0.8--1.1 & 87$\pm$3\\
		\hline                                             
	\end{tabular}
	\tablefoot{Both filters of ZIMPOL were set to the Very Broad Band (\textit{VBB}) filter covering a wide wavelength regime from \textit{R}- to \textit{I}-band. The following coronagraphs were used: N\_ALC\_YJH\_SDIT for IRDIS/IFS and V\_CLC\_S\_WF for ZIMPOL. DIT stands for the detector integration time and NDIT corresponds to the number of frames in the sequence. PC indicates the number of polarimetric cycles. The Strehl is calculated for the \textit{H}-band. This leads to a Strehl of $\sim$43\% for the ZIMPOL data at 0.65\,$\mu$m.}
\end{table*}

Observations of T~Cha were performed during the nights of 30~May~2015, 19~February~2016, and 31~March~2016 with several sub-systems of the high-contrast imager SPHERE equipped with an extreme adaptive optics system (SAXO, \citealt{fusco2006}, \citeyear{fusco2014}) and mounted on the VLT at Cerro Paranal, Chile. All observations were part of the SPHERE consortium guaranteed time program under IDs 095.C-0298(B) and 096.C-0248(B/C). The Infra-Red Dual-beam Imager and Spectrograph (IRDIS, \citealt{dohlen2008}) and the Zurich IMaging POLarimeter (ZIMPOL, \citealt{thalmann2008,schmid2012}) were used in Dual-band Polarimetric Imaging (DPI) mode (\citealt{langlois2014}) and in field stabilized (P2) mode, respectively. In addition, data were taken simultaneously with IRDIS in dual-band imaging (DBI; \citealt{vigan2010}) mode and the Integral Field Spectrograph (IFS, \citealt{claudi2008}). In this IRDIFS mode, IRDIS is operated in the filter pair \textit{H2H3} ($1.593$\,$\mu$m and $1.667\,\mu$m) and IFS in \textit{YJ} ($0.95$ -- $1.35\,\mu$m) mode. Table \ref{tab:obs} summarizes the observations and instrumental setups for each instrument. The Strehl ratio estimation (provided by SPARTA files) is based on an extrapolation of the phase variance deduced from the reconstruction of SAXO open-loop data using a deformable mirror, tip-tilt voltages, and wavefront sensor closed-loop data (\citealt{fusco2004}). The observing conditions and the different data reduction methods for each data set taken by the various sub-systems are described in detail in Sects. \ref{subsec:obs_irddpi} -- \ref{subsec:obs_adi}.
	
\subsection{IRDIS-DPI (\textit{H}-band)}
\label{subsec:obs_irddpi}
The IRDIS-DPI observations of T~Cha were carried out on 19~February~2016 with the \textit{BB\_H} filter ($\lambda_{\mathrm{c}} = 1.625\,\mu$m) using an apodized pupil Lyot coronagraph with a mask diameter of $\sim$185\,mas (\citealt{soummer2005,boccaletti2008}). Dark and flat field calibration were obtained during the following day. Thirty polarimetric cycles were taken, consisting of one data cube for each of the four half wave plate (HWP) positions (0$^{\circ}$, 45$^{\circ}$, 22.5$^{\circ}$ and 67.5$^{\circ}$). Dedicated coronagraphic images were taken at the beginning and at the end of the science sequence to determine accurately the position of the star behind the coronagraph. For this calibration a periodic amplitude is applied to the deformable mirror, which produces four equidistant, crosswise satellite spots of the stellar PSF outside of the coronagraph. The data were reduced following the prescriptions of \citet{avenhaus2014} and \citet{ginski2016}, who consider the radial Stokes formalism. The first step consists of standard calibration routines, including dark-frame subtraction, flat-fielding and bad-pixel correction. These images are split into two individual frames representing the left and right sides (parallel and perpendicular polarized beams, respectively), and the precise position of the central star is measured using the star center calibration frames on both image sides separately. Then, the right side of the image is shifted and subtracted from the left side. To obtain clean Stokes $Q$ and $U$ images, that is, to correct for instrumental polarization downstream of the HWP's position in the optical path, $Q^{+}$ and $Q^{-}$ (0$^{\circ}$ and 45$^{\circ}$), and $U^{+}$ and $U^{-}$ (22.5$^{\circ}$ and 67.5$^{\circ}$) are subtracted, respectively. However, there might be still an instrumental polarization left upstream of the HWP in the final $Q$ and $U$ images, which is assumed to be proportional to the total intensity image as shown in \citet{canovas2011}. To obtain this residual instrumental signal, the azimuthal Stokes components are computed from (cf. \citealt{schmid2006})

\begin{align}
	Q_{\phi} &= +Q \cos\,2\phi + U \sin\,2\phi \,, \\
	U_{\phi} &= -Q \sin\,2\phi + U \cos\,2\phi \,.
	\label{eq:stokesphi}
\end{align}

\noindent The azimuth $\phi$ is defined with respect to the stellar position (x$_{0}$,y$_{0}$) as

\begin{equation}
	\phi = \mathrm{arctan} \frac{x-x_0}{y-y_0}\,.
	\label{eq:azimuth}
\end{equation}

As shown by \citet{canovas2015}, the signal in the $U_{\phi}$ frame should be small for a centrally illuminated symmetrical disk. We thus determined the scaling factor for our second instrumental polarization correction such that the (absolute) signal in an annulus around the central star in the $U_{\phi}$ frame is minimized. We then subtract the scaled Stokes $I$ frame from the $Q$ and $U$ frame and use these final corrected frames to create the $Q_{\phi}$ and $U_{\phi}$ images displayed in Fig.~\ref{fig:pdiobs}.\\

To cross-check the IRDIS-DPI results, and especially to test the reliability of the $U_{\phi}$ minimization technique in the context of an inclined disk, we additionally perform an alternative reduction procedure. This includes a proper polarimetric calibration using a Mueller matrix model, whose details will be presented in van Holstein et al. (in prep.) and de Boer et al. (in prep.). A short explanation of the method and the corresponding results for T~Cha can be found in Sect.~\ref{subsec:res_pi} and Fig.~\ref{fig:pdiobs_corr}.

\subsection{ZIMPOL P2 (\textit{VBB} filter)}
\label{subsec:obs_zimpol}

T~Cha was observed during the night of 31~March~2016 with the SlowPolarimetry detector mode of ZIMPOL using the very broad band filter (\textit{VBB}). The \textit{VBB} filter covers a wide wavelength range from \textit{R}- to \textit{I}-band (0.55--0.87\,$\mu$m). These observations were also obtained with an apodized Lyot coronagraph (mask diameter of $\sim$185\,mas).\\

\begin{figure*}
	\centering
	\includegraphics[width=1.0\textwidth]{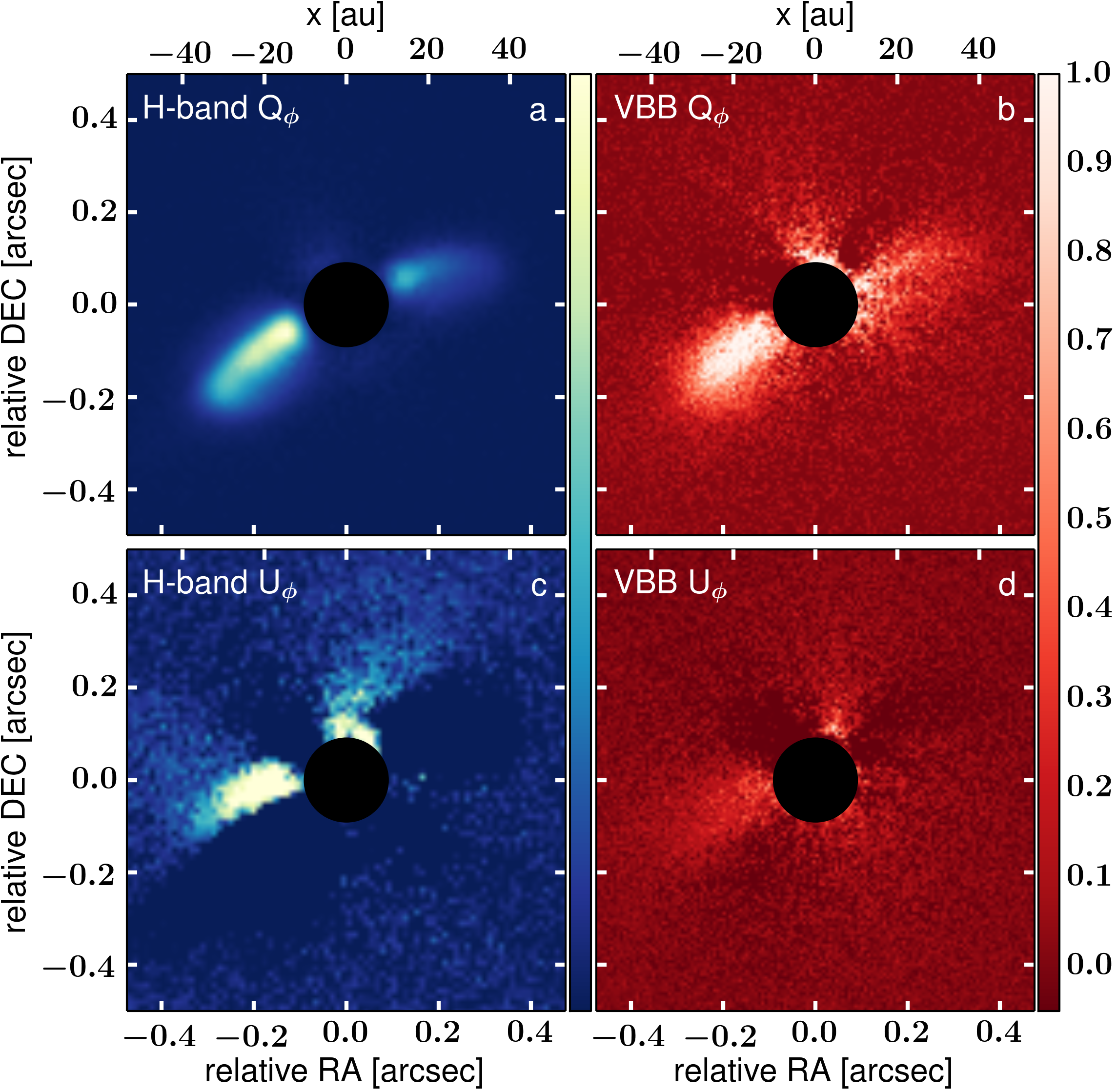}
	\caption{IRDIS-DPI \textit{H}-band and ZIMPOL P2 \textit{VBB}-filter $Q_{\phi}$ (top row) and $U_{\phi}$ (bottom row) images. North is up, east is left. All images are normalized to the highest disk brightness. The dynamical range for the color scaling is the same for the two images of the top (1000) and bottom row (20), respectively. An apodized Lyot coronagraph with a mask diameter of $\sim$185\,mas was used. The inner 0\farcs18 are masked, represented by the black circular area. Negative values of $U_{\phi}$ are saturated at dark blue and dark red color, respectively.}
	\label{fig:pdiobs}
\end{figure*}

The ZIMPOL data were reduced following mostly the same strategy as described for the IRDIS data in the previous section. The main difference between the two data sets is the different structure of the ZIMPOL data. In ZIMPOL the two perpendicular polarization directions for each HWP position are recorded quasi-simultaneously on the same detector pixels. For a more detailed description of the instrument and the specialized data reduction steps involved we refer to \citet{thalmann2008} and \citet{schmid2012}. We process both ZIMPOL detector images independently and only combine the images after the final data reduction to increase the S/N. We first bias subtract and flat field the individual frames. We then extract the two perpendicular polarization directions from the interlaced rows in each frame, resulting in two $1024 \times 512$ pixel images per original frame.
We then correct for charge-shifting artifacts by always combining two consecutive frames of the observation sequence. To create quadratic images we then bin each image by a factor of two along the x-axis. Finally we subtract the two perpendicular polarization directions from each other to get $Q^+$, $Q^-$, $U^+$ and $U^-$ frames (depending on the HWP position). These are then combined to create the final $Q$ and $U$ frames identical to the IRDIS reduction. In a last step we again calculate the azimuthal Stokes components $Q_\phi$ and $U_\phi$ and employ the instrumental polarization correction from \citet{canovas2011}. The resulting images (after combination of both ZIMPOL images) are also displayed in Fig.\ref{fig:pdiobs}.

\subsection{IRDIS-DBI (\textit{H2H3}-bands) and IFS (\textit{YJ}-band)}
\label{subsec:obs_adi}

T~Cha IRDIFS observations where obtained during the night of 30~May~2015 as part of the SpHere INfrared survey for Exoplanets (SHINE; Chauvin et al., in prep.) using the SHINE standard setup: pupil-stabilized images with IFS operating in \textit{YJ} mode (39 channels between 0.95 and 1.35\,$\mu$m) and IRDIS working in DBI mode using the \textit{H2H3} filter pair ($\lambda_{H2}=1.593\,\mu$m; $\lambda_{H3}=1.667\,\mu$m). This observing strategy allows for performing ADI (\citealt{marois2006}) in order to reach high contrast. The spectral resolution of IFS \textit{YJ} data amounts to R$\sim$50. The observations lasted about 6100 seconds with a field rotation of $\sim$28\degree. Since the target is located far to the south, obtaining a good rotation is challenging. The unstable weather conditions (Differential Image Motion Monitor (DIMM) seeing varied from 0\farcs5 to 0\farcs85 and clouds passing by) caused flux variations of up to one order of magnitude during the sequence.\\

The basic steps of the first IRDIS data reduction consist of flat-field and bad-pixel correction, cosmic ray detection and correction, and sky subtraction. Because of the variable atmospheric conditions during the observations, a very strict frame selection is applied at the end of the basic reduction and eventually only 42 out of 96 frames were used. This corresponds to selecting frames with Strehl ratio larger than  $\sim$25\%. The modeling and subtraction of the speckles follows the MPIA-PCA pipeline from Andr\'{e} M\"{u}ller. This is based on a principal component analysis (PCA) after \citet{absil2013}, which uses the Karhunen-Lo\'eve Image Projection (KLIP) algorithm from \citet{soummer2012}. We apply the following basic reduction steps: (1) Gaussian smoothing with half of the estimated FWHM; (2) normalization of the images based on the measured peak flux of the PSF images; (3) PCA and subtraction of the modeled noise; and (4) derotation and averaging of the images. Each frame is divided into annuli of the size of the estimated FWHM. The noise of a single frame and annulus is modeled by PCA using a fixed value of five modes, which is found to be the best value after several attempts. In addition, only frames for modeling the noise of the individual frame were selected where a minimum protection angle can be guaranteed. This means that with this adaptive approach the effect of self subtraction for an extended source is minimized and the S/N maximized.\\

To cross-check the IRDIS-DBI results, we perform a parallel reduction using the SPHERE Data Reduction and Handling pipeline (DRH, \citealt{pavlov2008}) implemented at the SPHERE Data Center. This includes dark and sky subtraction, bad-pixel removal, flat-field correction, anamorphism correction, and wavelength calibration. After these first steps the data were sorted according to their quality. Because of the difficult observing conditions, we use stringent frame selection (using 77 frames out of 96). This roughly corresponds to selecting frames with Strehl ratio greater than 15\% and leads to an average \textit{H-}band Strehl of $\sim$33\%. The location of the star is identified with the four symmetrical satellite spots generated from a waffle pattern on the deformable mirror. Then, to remove the stellar halo and to achieve high-contrast, the data were processed with the SpeCal pipeline developed for the SHINE survey (Galicher et al., in prep.); this implements a variety of ADI-based algorithms: Classical Angular Differential Imaging (cADI, \citealt{marois2006}), Template Locally Optimized Combination of Images (TLOCI, \citealt{marois2006}) and PCA (\citealt{soummer2012,amara2012}). In the following we discuss the results based on the TLOCI, and PCA images for the morphology and photometric analyses. Separate reductions were performed for the extraction of the disk. In particular, for the SpeCal PCA reduction a small number of PCA modes is used in order to enable optimal retrieval of the disk. The reduced numbers of modes, between two and four, are determined by maximizing the SNR inside a region delimited by the disk location. For the contrast curves, TLOCI images were considered because they provide the best compromise of contrast, stellar rejection, and throughput correction for the point source detection.\\ 

The data reduction for the IFS is performed using tools available at the SPHERE Data Center at IPAG following the procedure described in \citet{mesa2015} and in \citet{zurlo2014}. Using the SPHERE DRH software we apply the appropriate calibrations (dark, flat, spectral positions, wavelength calibration and instrument flat) to create a calibrated datacube composed of 39 images of different wavelengths for each frame obtained during the observations. Similar to the procedure used for IRDIS, in order to take into account the very variable weather conditions, we apply a frame selection resulting in 76 frames out of the original 96. A frame is considered as `bad' if the adaptive optics loop opens or the star exits the coronagraph region, causing an excess of light in the central part of the image. For each frame two central areas with 20 and 160 pixels per side are defined, for which the flux ratio is determined. Frames are rejected by an automated sorting if this ratio exceeds 130\% of the median value. The position of the star behind the coronagraph is estimated from images with four satellite spots, symmetric with respect to the central star taken just before and after the standard coronagraphic observations. Exploiting these images we are then able to define the re-scaling factor for images at different wavelengths to maintain the speckle pattern as stable as possible. Moreover, we are able to combine those images using the PCA algorithm from \cite{soummer2012} to implement both ADI and spectral differential imaging (SDI, \citealt{racine1999}) in order to remove the speckle noise.

\begin{figure*}
	\centering
	\centerline{
		\includegraphics[width=\textwidth]{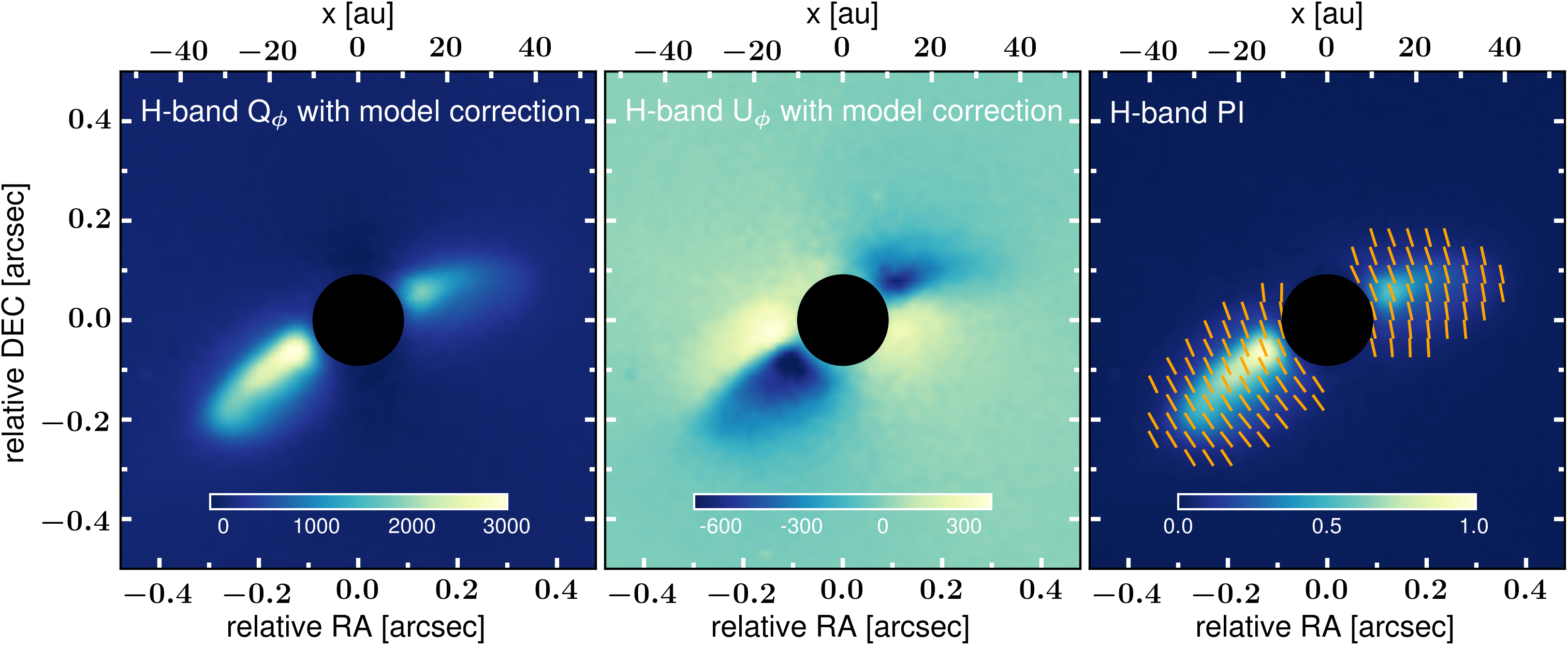}	
	}
	\caption{Mueller matrix model-corrected IRDIS-DPI $Q_{\phi}$ (left), $U_{\phi}$ (middle) and polarized intensity $PI=\sqrt{Q^2+U^2}$ (right) images. North is up, east is left. Note that the $Q_{\phi}$ and $U_{\phi}$ images are not normalized/saturated here on purpose to emphasize their partially negative signal. The inner 0\farcs18 are masked, represented by the black circular area. The orange stripes in the right image represent the angle of linear polarization (fixed length, not scaled with the degree of polarization).}
	\label{fig:pdiobs_corr}	 
\end{figure*}

\begin{figure*}
	\centering
	\centerline{
		\includegraphics[width=1.0\textwidth]{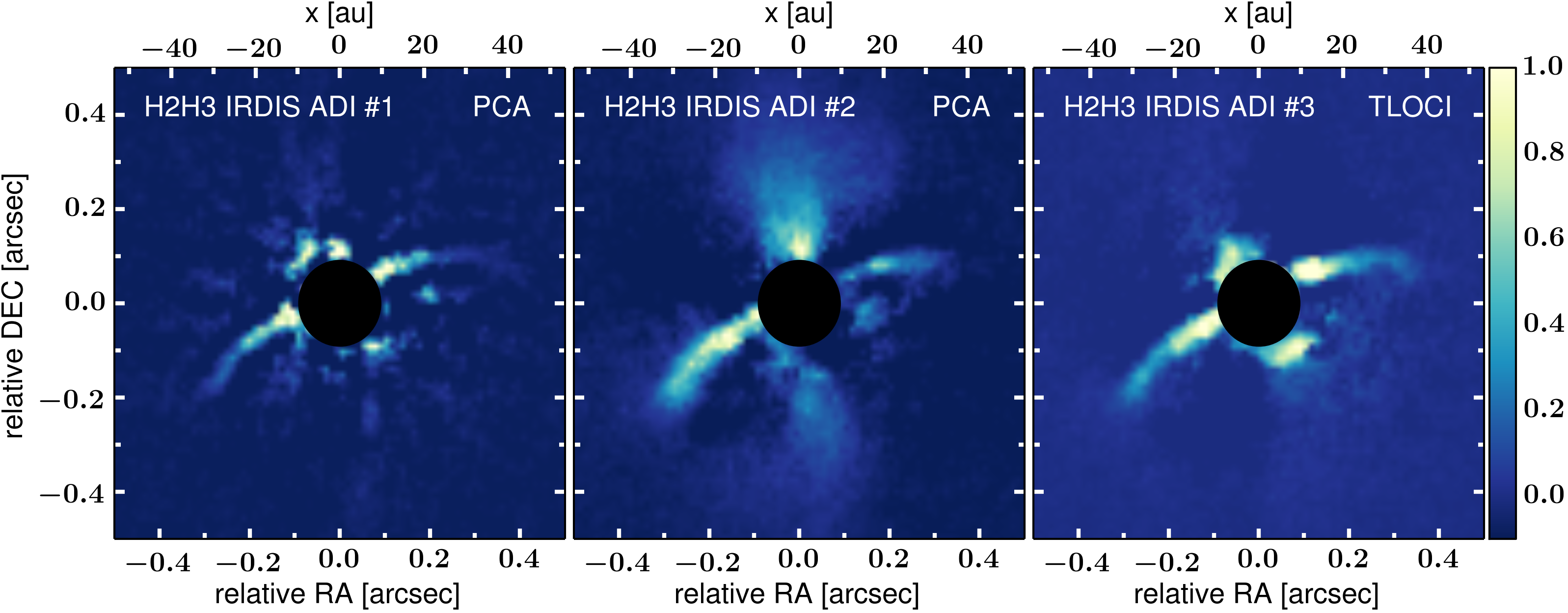}
	}
	\caption{IRDIS-ADI \textit{H2H3}-band images (mean across the wavelengths) based on three different reduction pipelines (\#1: MPIA-PCA, \#2: SpeCal-PCA and \#3: SpeCal-TLOCI). North is up, east is left. The images are normalized to the highest disk brightness and the color scales consider the same dynamical range. The inner 0\farcs18 region masked by the coronagraph is represented by the black circular area.}
	\label{fig:adiobs}	 
\end{figure*}

\section{Results}
\label{sec:results}

The disk of T~Cha is detected in all data sets presented in this study. The analysis of the disk geometry primarily focuses on the IRDIS-DPI and IRDIS-ADI images (see Sect. \ref{subsec:ana_disk}). Furthermore, the ADI images are used to search for point-source signals focusing on non-polarized companions, because this data set reaches a higher contrast (Sects. \ref{subsec:ana_ps} and \ref{subsec:ana_limits}).  

\subsection{Polarized intensity images}
\label{subsec:res_pi}

Figure \ref{fig:pdiobs} shows the reduced $Q_{\phi}$ and $U_{\phi}$ images of the IRDIS-DPI \textit{H}-band and ZIMPOL \textit{VBB} observations of T~Cha described in Sects. \ref{subsec:obs_irddpi} and \ref{subsec:obs_zimpol}. The dark central region corresponds to the area masked by the coronagraph. The disk is clearly detected in the IRDIS $Q_{\phi}$ image, which gives by far the best quality view of the outer disk structure and its rim in scattered light for T~Cha. Our SPHERE observations support a high disk inclination with respect to the line of sight, in agreement with the model by \citet{huelamo2015}. Scattered light is detected out to a projected distance of $\sim$0\farcs39 ($\sim$42\,au) from the central star concentrated in a bright arc with, however, a significant difference in brightness between the east and west sides (factor of $\sim$2.5). This is further discussed in Sects. \ref{subsubsec:ana_disk_model} and \ref{subsec:discu_asym}. In the reduced $U_{\phi}$ image there is some residual signal left, which has usually been interpreted as instrumental effects or as imperfect centering of the images. However, since the $U_{\phi}/Q_{\phi}$ peak-to-peak value amounts to 9\% and owing to the high inclination of the disk around T~Cha (69\,deg is determined from the total intensity image modeling, see Sect. \ref{subsubsec:ana_disk_model}), multiple scattering (e.g., \citealt{bastien1990,fischer1996,ageorges1996}), that is, scattering of already polarized light, in the inner disk might be the prime contributor to this signal. This is consistent with a theoretical study by \citet{canovas2015}, who found that even for moderate disk inclinations multiple scattering alone can produce significant non-azimuthal polarization above the noise level in the $U_{\phi}$ images. They showed that the $U_{\phi}/Q_{\phi}$ peak-to-peak value can even go up to 50\% for a disk inclination of 70\,deg depending on the mass and grain size distribution of the disk model. We note that the exact geometrical structure of the $U_{\phi}$ signal might be influenced by the reduction method described in Sect. \ref{subsec:obs_irddpi} (correction for the instrumental crosstalk by minimizing $U_{\phi}$). Therefore, we are not going to force our model to also fit the $U_{\phi}$ in addition to the $Q_{\phi}$. However, in order to prove that the $U_{\phi}$ signal is indeed real, we evaluate our IRDIS-DPI data with a newly developed reduction method, independent of the one presented in Sect. \ref{subsec:obs_irddpi}. By using the detailed Mueller matrix model of van Holstein et al. (in prep.) and de Boer et al. (in prep.) we correct our measurements for instrumental polarization effects. This model describes the complete optical path of SPHERE/IRDIS, i.e. telescope and instrument, and has been fully validated with measurements using SPHERE's internal source and observations of unpolarized standard stars (van Holstein et al., in prep.). The images of Stokes $Q$ and $U$ incident on the telescope are computed by setting up a system of equations describing every measurement of $Q$ and $U$ and solving it --- for every pixel individually --- using linear least-squares. The resulting $Q_{\phi}$ image (Fig.~\ref{fig:pdiobs_corr}, left panel) is very similar to the one from the first reduction (Fig.~\ref{fig:pdiobs}, top left panel). The new $U_{\phi}$ image (Fig.~\ref{fig:pdiobs_corr}, middle panel) has a higher accuracy than the one from the reduction that minimizes $U_\phi$ (Fig.~\ref{fig:pdiobs}, bottom left panel), in particular because no assumptions about the angle of linear polarization of the source are made to correct for the instrumental polarization. The $U_{\phi}$ image is cleaner and shows more symmetry in the sense that there is also a strong signal to the south-west. On the top right, the positive $U_\phi$ signal from Fig.\ref{fig:pdiobs}, bottom left, is not visible anymore. The right panel of Fig.~\ref{fig:pdiobs_corr} shows the polarized intensity overplotted with polarization vectors representing the angle of linear polarization. This strengthens that there is a clear departure from azimuthal polarization. For the model-corrected images the $U_{\phi}/Q_{\phi}$ peak-to-peak value increases from 9\% to 14\%, suggesting that some of the actual physical $U_{\phi}$ signal has been removed in the original reduction method due to the $U_{\phi}$ minimization procedure. A detailed comparison between different reduction methods and the specific influence on the left-over $U_{\phi}$ signal will be the topic of the two follow-up SPHERE papers.\\

\begin{figure}
	\centering
	\centerline{
		\includegraphics[width=\columnwidth]{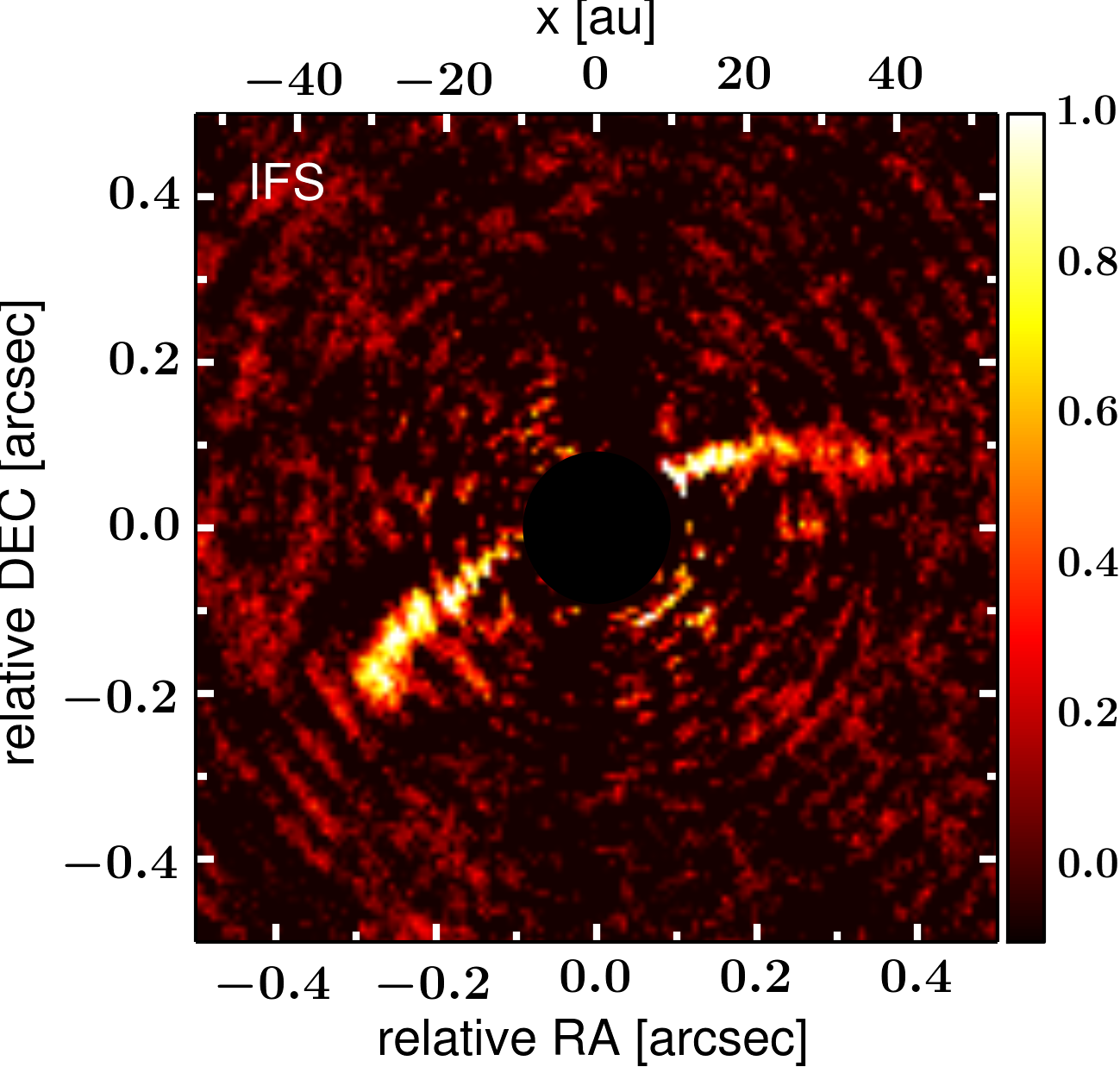}	
	}
	\caption{IFS image after PCA+SDI reduction with 100 modes: median across the entire wavelength range \textit{YJ}. North is up, east is toward the left; the image is normalized to the highest disk brightness and the color scale considers the same dynamical range as in Fig.\ref{fig:adiobs}. The inner 0\farcs18 region masked by the coronagraph is represented by the black circular area.}
	\label{fig:ifsobs}	 
\end{figure}

The optical images obtained with ZIMPOL (Fig.~\ref{fig:pdiobs}, right panel) corroborate the disk geometry, but the bad weather conditions and low Strehl ratio (43\% at 0.65\,$\mu$m) of this observation lead to a rather blurred structure. Again, positive and negative patterns (dark and bright color) alternate in the $U_{\phi}$ image, where these negative patterns are practically at the same location as in the IRDIS $U_{\phi}$ image.

\subsection{Total intensity images}
\label{subsec:res_intens}

In addition to the polarimetric images, the IRDIS-ADI \textit{H2H3} intensity images in Fig.~\ref{fig:adiobs} also clearly show the inclined disk around T~Cha. It even more strongly brings out the inner rim of the outer disk on the far side, visible as a faint arc below the coronagraph. The double-arch structure is a recurrent new form of features we have been detecting with high-contrast imaging instruments such as SPHERE (cf. \citealt{janson2016,garufi2016}). We note that because of the ADI processing this image may have been biased and is not a faithful representation of the true intensity and geometry. The residual signal northeast of the image center is likely due to stellar residuals. However, this signal is almost aligned with the near minor axis, so the possibility that it is real cannot be completely excluded. In fact, it could be the marginal detection of some material outward of the ring with high scattering efficiency. The brightness asymmetry between the west and east disk wings is as pronounced as in the polarimetric images, especially for reduction \#2. Figure \ref{fig:ifsobs} illustrates the median IFS image across the wavelength range from \textit{Y}- to \textit{J}-band. The disk is nicely resolved, confirming the disk geometry and surface brightness extension stated above.

\section{Analysis}
\label{sec:ana}

\subsection{Radiative transfer modeling of the disk}
\label{subsec:ana_disk}

We build a radiative transfer model for T~Cha aiming to reproduce the basic structure of its disk. We take earlier efforts (\citealt{olofsson2011, olofsson2013, huelamo2015}) as a starting point for independent three-dimensional (3D) radiative transfer calculations using the Monte Carlo code \textsc{RADMC-3D}\footnote{The \textsc{RADMC-3D} source code and more details are available online at http://www.ita.uni-heidelberg.de/$\sim$dullemond/software/radmc-3d/.}\citep{dullemond2012}. We aim to complement the current understanding of the disk geometry by also taking into account our new SPHERE data. \textsc{RADMC-3D} is used to calculate the thermal structure of the dust disk and ray-traced synthetic scattered light images in the NIR. The polarization of \textsc{RADMC-3D} was investigated by \cite{kataoka2015}, who performed a benchmark test against the numerical models presented in \cite{pinte2009}.

\subsubsection{Modeling approach}
\label{subsubsec:ana_disk_methods}

The disk around T~Cha is parametrized using constraints obtained from previous analyses of data sets (cf., \citealt{olofsson2011, olofsson2013, huelamo2015}) and from the new SPHERE observations presented in this paper. We assume the disk to be composed of two spatially separated zones with an inner (r$_{\mathrm{in}}$) and outer radius (r$_{\mathrm{out}}$) each, a narrow inner disk close to the central star that is responsible for the NIR excess and a more extended outer disk. Inner and outer disks are assumed to be coplanar, since there is no significant evidence for a misaligned inner disk, which would cast shadows onto the outer disk (cf. the cases of HD142527, \citealt{marino2015} and HD100453, \citealt{benisty2017}). The surface density structure is defined by a power-law profile and an exponential taper at the outer edge of the outer component (e.g., \citealt{hughes2008}),

\begin{equation}
	\Sigma(r) = \Sigma_0 \left( \frac{r}{r_{\mathrm{c}}} \right)^{-\delta} \exp \left[ -\left( \frac{r}{r_{\mathrm{c}}} \right)^{2-\delta} \right]\,,
	\label{eq:density}
\end{equation}

\noindent where $r_{\mathrm{c}}$ corresponds to a characteristic radius and $\delta$ denotes the surface density index. For the sake of simplicity, we assume a uniform distribution along the azimuth in our model and concentrate on the radial disk structure. The disk scale height is parameterized radially as $H(r)=H_{0}\,(r/r_{0})^{\beta}$, where $H_{0}$ is the scale height at a reference radius $r_{0}$ and $\beta$ is the flaring index. The vertical density distribution follows a Gaussian profile, so that the dust volume density is given by

\begin{equation}
	\rho(R,\varphi,z) = \frac{\Sigma(R)}{\sqrt{2\,\pi}\,H(R)}\,\exp \left( -\frac{z^2}{2\,H^2(R)} \right)\,,	
	\label{eq:density3d}
\end{equation}

\noindent where the spherical coordinates $R$ and $z$ can be converted into cylindrical coordinates with $R = r\,\sin(\theta)$ and $z = r\,\cos(\theta)$, where $\theta$ is the polar angle. We consider a power-law grain size distribution with an index $p=-3.5$, dn(a) $\propto$ a$^{p}$\,da between a minimum (a$_{\mathrm{min}}$) and maximum grain size (a$_{\mathrm{max}}$). During our modeling process we use different values for the parameters a$_{\mathrm{min}}$ and a$_{\mathrm{max}}$, where two distributions are eventually find to give an equally good match for the total intensity image (cf. Sect. \ref{subsubsec:ana_disk_model}). The dust is assumed to be a mixture made of silicates (\citealt{draine2003b}), carbon (\citealt{zubko1996}), and water ice (\citealt{warren2008}) with fractional abundances of 7\%, 21\%, and 42\%, consistent with \citet{ricci2010}. The remaining 30\% is vacuum. The opacity of the mixture is determined by means of the Bruggeman mixing formula. The absorption and scattering opacities, $\kappa_{\mathrm{scat}}$ and $\kappa_{\mathrm{abs}}$, as well as the scattering matrix elements $Z_{ij}$ are calculated for spherical, compact dust grains with Mie theory considering the BHMIE code of \citet{bohren1983}.\\

\renewcommand{\arraystretch}{1.2}
\begin{table}
	\caption{Overview of the best \textsc{RADMC-3D} model parameters}              
	\label{tab:rt}      
	\centering
	\begin{tabular}{lcc}          
		\hline\hline                       
		Parameter & Inner disk & Outer disk\\
		\hline
		r$_{\mathrm{in}}$ [au]$^{\ast}$ & 0.07$^{a}$ & 30\\
		r$_{\mathrm{out}}$ [au]$^{\ast}$ & 0.11$^{a}$ & 60\\
		r$_{\mathrm{c}}$ [au] & -- & 50$^{b}$\\
		M$_{\mathrm{dust}}$ [M$_{\odot}$]& $2 \cdot 10^{-11}$\,$^{a}$ & $9 \cdot 10^{-5}$\,$^{b}$\\
		$\delta$ & 1 & 1\\
		H$_{0}$/r$_{0}$ & 0.02/0.1$^{a}$ & 4/50$^{b}$\\
		$\beta$ & 1 & 1\\
		\hline
		$\lbrace$a$_{\mathrm{min}}$,a$_{\mathrm{max}}\rbrace$ [$\mu$m]$^{\ast}$ & $\lbrace$0.01,1000$\rbrace$ & $\lbrace$0.01,1000$\rbrace$\\
		& $\sim$10 & $\sim$10\\
		p & -3.5 & -3.5\\
		\hline
		incl [deg]$^{\ast}$ & 69 & 69\\
		PA [deg]$^{\ast}$ & 114 & 114\\
		\hline                                             
	\end{tabular}
	\tablefoot{$\delta$ denotes the exponent of the surface density power-law and $\beta$ corresponds to the disk flaring index. For the radiation source we take the following star parameters: T$_{\mathrm{eff}}=5400\,K$, M$=1.5\,M_{\odot}$, R$=1.3\,R_{\odot}$, where the star is assumed to be spherical. All parameters marked with an asterisk symbol ($^{\ast}$) were varied during the radiative transfer modeling. The grain size distributions as well as inclination and PA were taken to be the same for the inner and outer disk. References: $^{a}$\,\citet{olofsson2013}; $^{b}$\,\citet{huelamo2015}.}
\end{table}

The radiative transfer calculations start with computing the dust temperature consistently by means of a thermal Monte Carlo simulation using $10^{7}$ photon packages\footnote{Each single package actually represents many photons at once assuming that these photons follow the same path.}. Hence, an equilibrium dust temperature is calculated considering the star as the source of luminosity. The main inputs for the radiative transfer modeling are the dust density structure from Eq. \ref{eq:density3d} and the dust opacities. Full non-isotropic scattering calculations are performed that take multiple scattering and polarization into account. To compare with the observations, synthetic Stokes $I$ intensity images, and Stokes $Q$ and $U$ polarized intensity images are produced at \textit{H}-band (1.6\,$\mu$m) using $10^{8}$ photon packages. These theoretical images are then convolved with a Gaussian PSF with a FWHM of $0\farcs04$ assuming the object to be at 107\,pc. Moreover, the synthetic total intensity images are run through the MPIA-PCA and SpeCal-PCA processing described in Sect. \ref{subsec:obs_adi} to have a proper comparison. The polarimetric Stokes $Q$ and $U$ images are eventually converted into their azimuthal counterparts $Q_{\phi}$ and $U_{\phi}$. All synthetic images are normalized to the highest disk surface brightness and displayed using the same dynamical range as for the observational data. The coronagraph used in our IRDIS observations is mimicked by masking the inner 0\farcs18 of the disk (19.3\,au at 107\,pc distance).

\subsubsection{Best model}
\label{subsubsec:ana_disk_model}

\begin{figure*}
	\centering
	\centerline{
		\includegraphics[width=\textwidth]{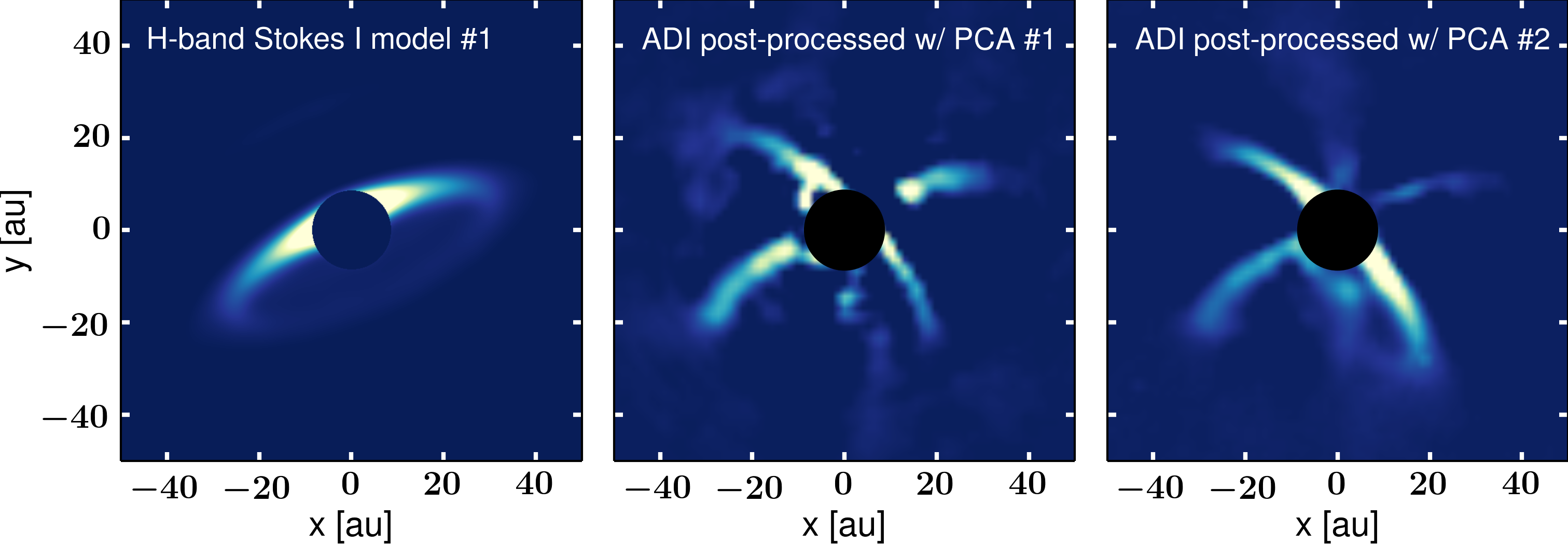}
	}
	\caption{Synthetic total intensity images from our radiative transfer model \#1 considering a MRN-like power-law grain size distribution with $a_{\mathrm{min}}=0.01\,\mu$m and $a_{\mathrm{max}}=1000\,\mu$m. The left panel shows the theoretical Stokes~I image convolved with a Gaussian PSF with FWHM of 0\farcs04 (at 107\,pc distance). The middle and right panels show the theoretical model image at 70$^{\circ}$ processed together with the raw DBI data by the different PCA methods as described in Sect. \ref{subsec:obs_adi}. The central 0\farcs18 of the image are masked to mimic the effect of the coronagraph on the observations. The units are arbitrary, but the dynamical range of the color bar is taken the same as in Fig.\ref{fig:adiobs}.}
	\label{fig:rt_adi_mod1}
\end{figure*}

\begin{figure*}
	\centering
	\centerline{
		\includegraphics[width=\textwidth]{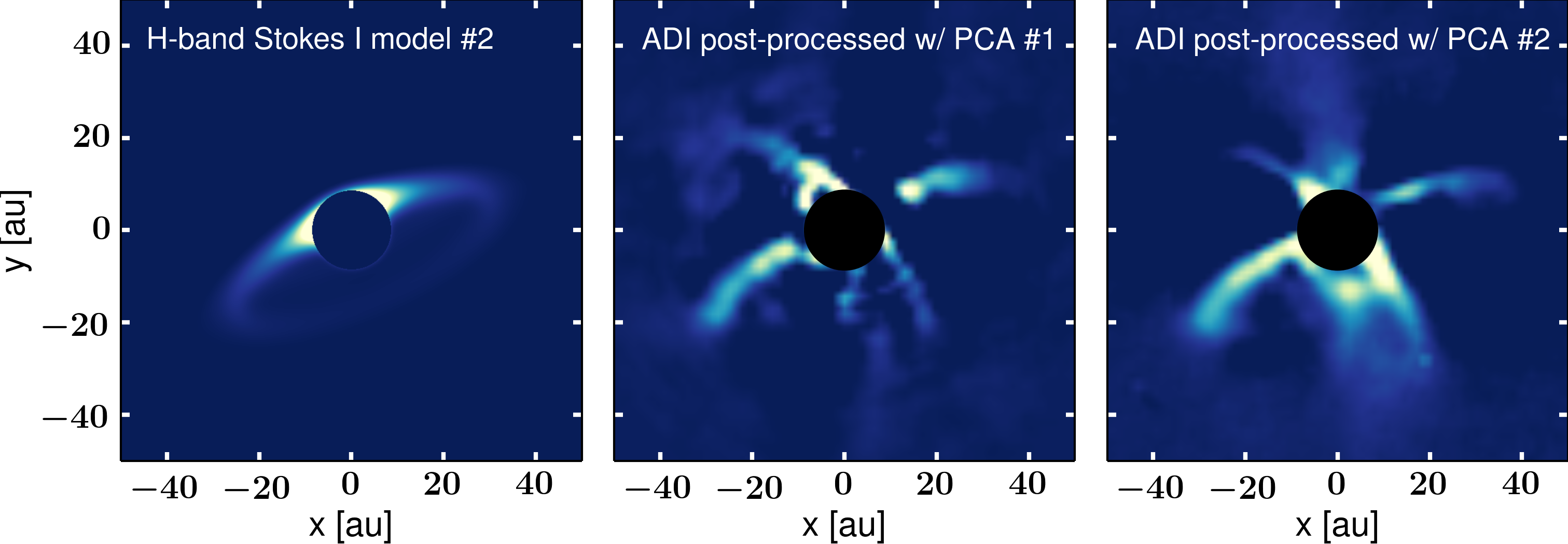}
	}
	\caption{Synthetic total intensity images from our radiative transfer model \#2 considering a grain size distribution narrowly peaked around $\sim$10\,$\mu$m. The layout and color scale is identical to Fig.~\ref{fig:rt_adi_mod1}.}
	\label{fig:rt_adi_mod2}
\end{figure*}

The inner disk geometry parameters are adopted from \citet{olofsson2013} and are kept fixed in the modeling process. By adjusting the parameters from \citet{huelamo2015} we generate the outer disk and run a grid of models exploring a pre-defined parameter space for $r_{\mathrm{in}}$, $r_{\mathrm{out}}$, $a_{\mathrm{min}}$, $a_{\mathrm{max}}$, incl and PA. The fiducial model is defined by the set of parameters that causes a minimization in the residuals between observations and model within the paramater ranges set. For this determination the images of both, data and models, are normalized to the highest flux value outside of the coronagraph. There is, however, no automatic fitting routine since computing tens of thousands of 3D models for the T~Cha system is computationally far too expensive. The best parameters are summarized in Table \ref{tab:rt}. The modeling approach taking into account the new high-contrast SPHERE images allow us to better constrain the position of the inner rim of the outer disk, which we find to be at a significantly larger radius of $\sim$30\,au ($\sim$0\farcs28) compared to earlier work. Hence, the cavity size between the inner and outer disk is correspondingly larger than previously thought (\citealt{olofsson2013}). Furthermore, the polarimetric measurements provide us better estimates of the grain sizes.\\

\begin{figure*}
	\centering
	\centerline{
		\includegraphics[height=0.24\textwidth]{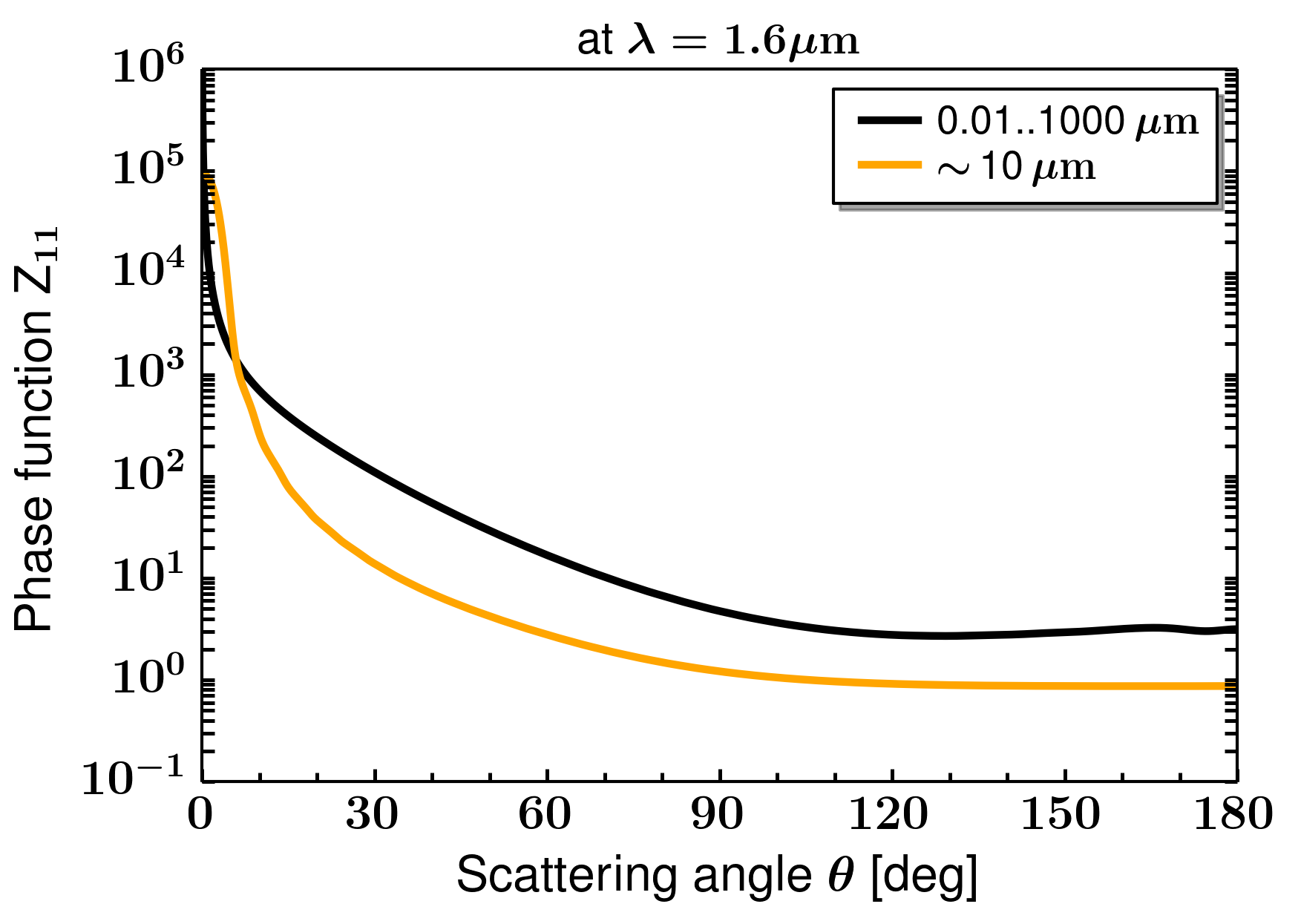}
		\includegraphics[height=0.24\textwidth]{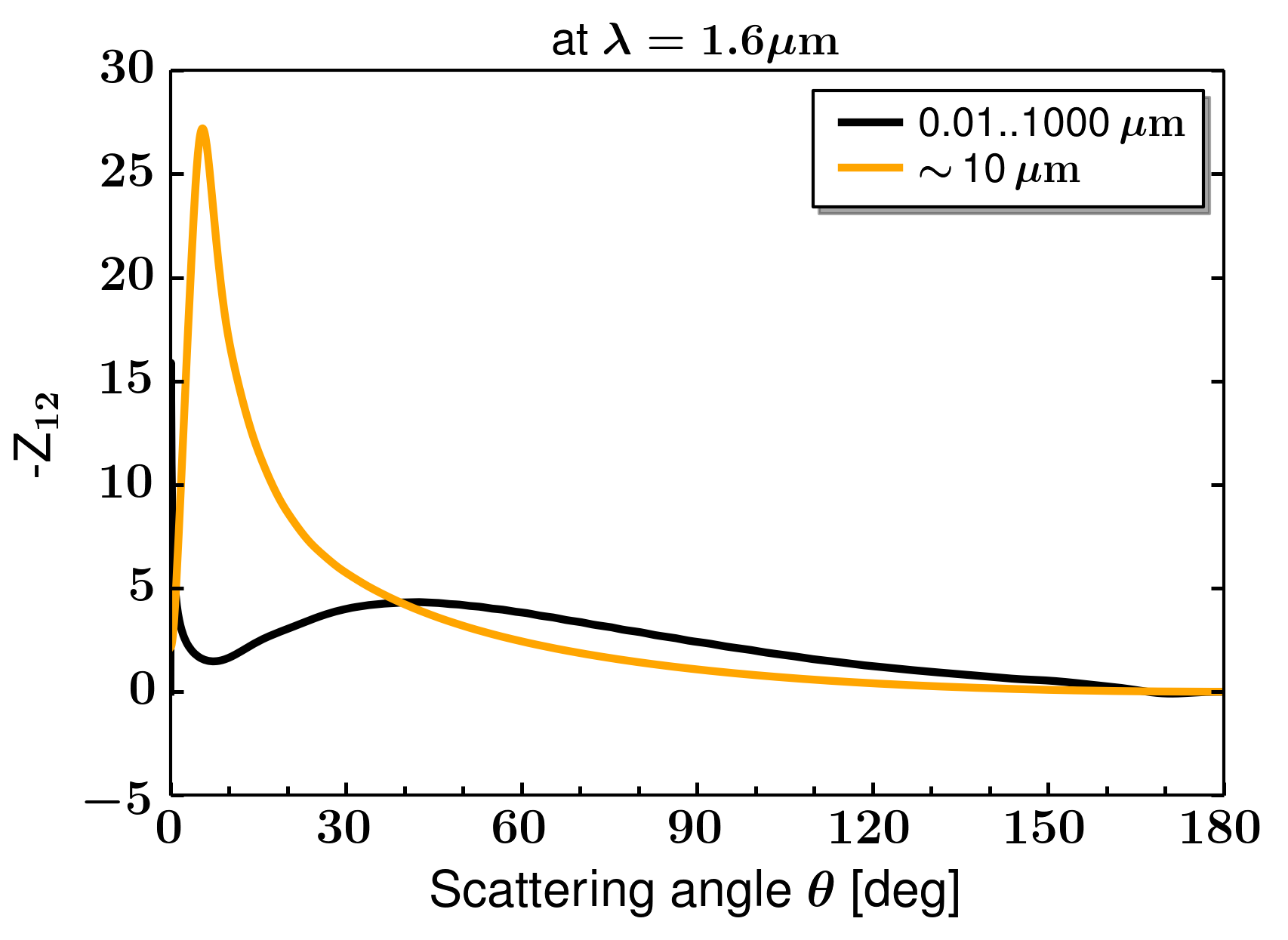}
		\includegraphics[height=0.24\textwidth]{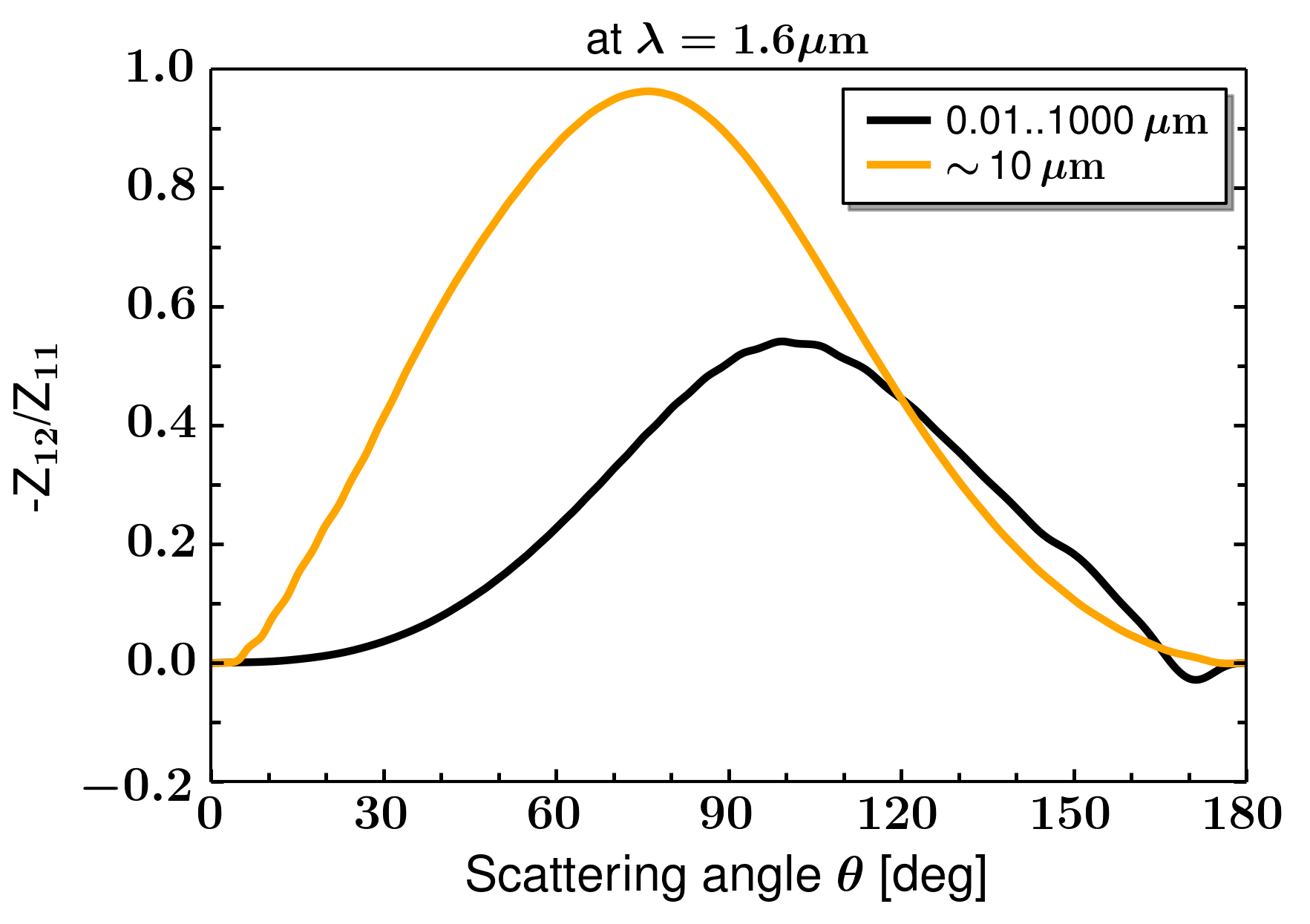}		
	}
	\caption{Phase functions $Z_{\mathrm{11}}$, $-Z_{\mathrm{12}}$ and degree of polarization $-Z_{\mathrm{12}}/Z_{\mathrm{11}}$ of the dust grains dependent on the scattering angle $\theta$ and calculated at $\lambda=1.6\,\mu$m. Model \#1 is represented by the black line, model \#2 by the orange curve.}
	\label{fig:matrix}	 
\end{figure*}

\subsubsection*{Synthetic total intensity images}
Figure \ref{fig:rt_adi_mod1}, left panel, shows the synthetic Stokes $I$ image at \textit{H}-band from the first of our two radiative transfer models. It is produced at a disk position angle of $\mathrm{PA}=114^{\circ}$ and an inclination angle of $\mathrm{i}=69^{\circ}$, which is similar to those values derived in \citet{huelamo2015}. Our disk model gives a qualitatively good match with the IRDIS total intensity images from Fig.\ref{fig:adiobs}. The bright arc as the dominant source of scattered light is well reproduced and corresponds to forward scattered light from the near side of the inclined disk. The ADI images may, however, be significantly altered by the software processing, which was already shown by \citet{garufi2016} for the case of HD100546. This ADI bias is especially important for T~Cha, since the self-subtraction is strong due to the small field rotation and high inclination. Thus, we apply the ADI processing routines to the model image. To do so we process the model image rotated by 70\,deg with the raw data considering the same PCA parameters. The middle and right panels of Fig.~\ref{fig:rt_adi_mod1} show the resulting post-processed images depending on the PCA reduction method. The ADI procedure damps the signal of the backside of the disk and introduces a brightness asymmetry along the disk surface. Thus, an original azimuthally symmetric feature can be seen as an asymmetric double-wing structure for a specific disk geometry and orientation. We note here that we additionally favor a physical reason for this asymmetry, since this is also seen in the polarimetric images (cf. Sect. \ref{subsec:discu_asym}). The ADI processed model image supports that the geometrical parameters used in our model, in particular the gap size, reproduce the observations nicely.\\ 

The fraction of star light scattered off the disk surface layer towards the observer depends on the disk properties (e.g., mass and scale height), but also on dust grain properties that determine the phase function. Dust grains, which are large compared to the wavelength, have strongly forward peaking scattering phase function, while small grains scatter photons almost isotropically. When keeping the minimum dust grain size fixed at 0.01\,$\mu$m, a maximum grain size of at least $100\,\mu$m is requested to match the observations. This serves to reduce the influence of the small grains  that are in the Rayleigh limit and absorb radiation much more efficiently than they scatter it. Except for very turbulent disks, one would, however, expect very large grains (> 10\,$\mu$m) to settle below the scattering surface. The need for large grains in the disk surface can be avoided by removing the smallest grains. Hence, the minimum and maximum values for the dust grain size distribution in our model are somehow degenerate. An equally good image, that also achieves the desired brightness contrast of the arc with respect to the disk backside, is obtained by using a narrow distribution around $10\,\mu$m. Grains of about ten microns in size are strong forward scatterers in the \textit{H}-band. If even larger particles were primarily present, the forward scattering efficiency would be too strong, and the brightness of the disk's far sides would be too faint. The corresponding synthetic intensity images for the second model and their appearance after the ADI post processing with PCA can be found in Fig.~\ref{fig:rt_adi_mod2}.

\subsubsection*{Synthetic polarimetric images}
Our results so far demonstrate that we find a quite good model to match the disk geometry of T~Cha. The goal is, however, to also analyze the grain properties compatible with the polarimetric data. For scattering in the Rayleigh and Mie regime, that is, for grains with sizes smaller than or approximately equal the wavelength ($2 \pi a \lesssim \lambda$), maximum polarization is expected along a scattering angle of 90\,deg. The phase function $Z_{\mathrm{11}}$, the scattering matrix element $-Z_{\mathrm{12}}$ and the degree of polarization $-Z_{\mathrm{12}}/Z_{\mathrm{11}}$ of the dust grains used in our radiative transfer models are shown as a function of the scattering angle $\theta$ in Fig.~\ref{fig:matrix}. Comparing those quantities for both models allows us to rule out the first model covering a wide range of grain sizes from 0.01 to 1000\,$\mu$m. This, rather, produces maxima in polarized intensity along the semi-major axis (see Appendix \ref{app:model1}), which is clearly not observed in the SPHERE PDI data from Fig.~\ref{fig:pdiobs}. Although one can recognize an extreme forward peak in $-Z_{\mathrm{12}}$, the resulting peak in polarized intensity is hidden behind the coronagraph. For our second model with grains of $\sim$10\,$\mu$m the phase function is also dominated by small-angle scattering as seen in the $Z_{\mathrm{11}}$ plot, but the $-Z_{\mathrm{12}}$ curve has a strong peak at small angles of $\sim$10$^{\circ}$. This leads to the spatial shift of brightness maxima away from the semi-major axis (i.e., scattering at 90\,deg), meaning that the maximum polarized intensity occurs at the forward scattering position. This is in good agreement with our polarimetric SPHERE observations.\\

\begin{figure*}
	\centering
	\includegraphics[width=1.0\textwidth]{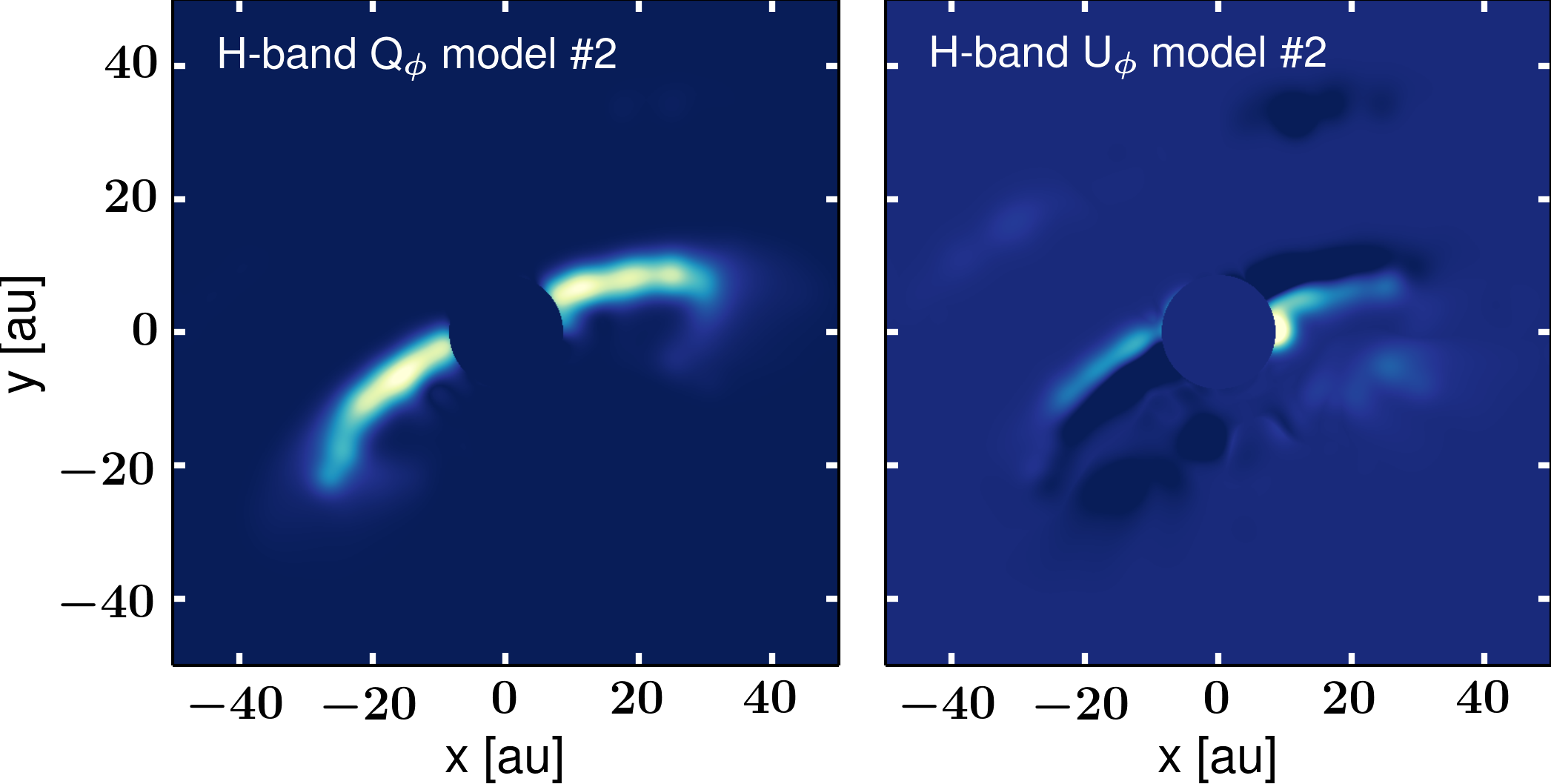}
	\caption{Synthetic $Q_{\phi}$ (left) and $U_{\phi}$ (right) images at \textit{H}-band. They are convolved with a Gaussian PSF with FWHM of 0\farcs04 (at 107\,pc distance). The color scale is arbitrary, the dynamical range is similar as in Fig.\ref{fig:pdiobs}. Negative values of $U_{\phi}$ are saturated at dark blue color.}
	\label{fig:rt_quphi}
\end{figure*}

Figure \ref{fig:rt_quphi} shows the synthetic $Q_{\phi}$ and $U_{\phi}$ images at \textit{H}-band for the second model, with a disk position angle of $\mathrm{PA}=114^{\circ}$ and an inclination of $\mathrm{i}=69^{\circ}$; both determined from the fit to the total intensity image. The $Q_{\phi}$ image is dominated by large positive signal, which is consistent with forward scattering from the close edge of the disk. The small-scale brightness blobs could be due to self-scattering of thermal emission or the result of multiple scattering treatment in the radiative transfer calculations. Monte Carlo noise can be ruled out as the source of these features since the best models were also run with a higher number of photon packages ($10^9$) for testing, confirming that our calculations are converged. Similar to the observed $U_{\phi}$ image, the $U_{\phi}$ model image shows an alternation of positive (white) and negative (dark blue) signal, although the exact geometry appears different. The extension of the south-east lobe with negative signal is comparable to that in the observational image in Figs.~\ref{fig:pdiobs} (bottom left panel) and \ref{fig:pdiobs_corr}. The positive signal is a bit less pronounced in our calculated model. Since the $U_{\phi}$ signal in the observational image can be substantially influenced by noise, instrumental effects, and the data reduction procedure, which is not included in our modeling, such a deviation was to be expected. The $U_{\phi}$/$Q_{\phi}$ peak-to-peak value for the best model is about 15\%, which is still in very good agreement with the observations (9\% and 14\%), but lower than calculated in the study by \citet{canovas2015} on non-azimuthal linear polarization. For their models and in our \textsc{RADMC-3D} calculations we consider a full treatment of polarized scattering off randomly oriented particles. Due to the absence of any instrumental influence on the polarization, the $U_{\phi}$ signal visible in the model images should be primarily connected to multiple scattering events happening in the disk. However, the contribution of multiple scattering strongly depends on the disk inclination, the grain population and the mass of the disk. In \citet{canovas2015} the signal in $U_{\phi}$ reaches up to 50\% of the $Q_{\phi}$, but only for an inclination of 70\,deg, a grain size distribution with a$_{\mathrm{min,max}}=(5, 1000)\,\mu\mathrm{m}$, and a disk significantly more massive than assumed for our T~Cha model. A higher disk mass produces more scattering events as simply more scattering particles are available. Furthermore, the higher scattering efficiency of the grains relative to their absorption efficiency results in stronger multiple scattering signature in the models of \citet{canovas2015}. These effects can explain the discrepancy to our $U_{\phi}$/$Q_{\phi}$ peak-to-peak value of only 15\%.

\begin{figure*}
	\centering
	\includegraphics[width=1.0\textwidth]{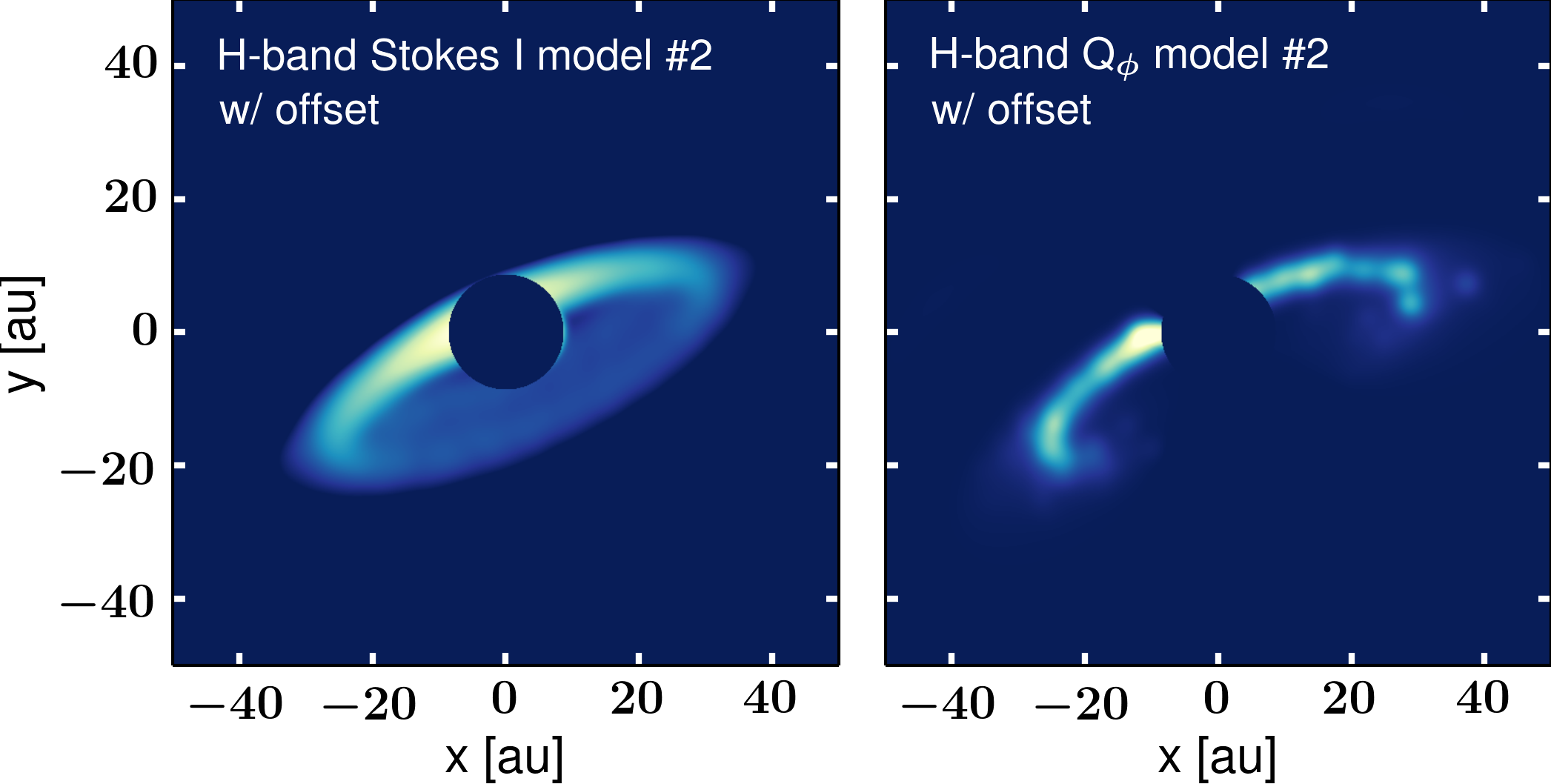}
	\caption{Synthetic Stokes $I$ (left) and $Q_{\phi}$ (right) images at \textit{H}-band of model \#2, where the star is slightly offset from its original central position along the semi-major axis. The images are convolved with a Gaussian PSF with FWHM of 0\farcs04 (at 107\,pc distance). The color scales are identical to Figs. \ref{fig:rt_adi_mod1} and \ref{fig:rt_quphi}, respectively.}
	\label{fig:rt_off}
\end{figure*}

\subsubsection*{East-west brightness asymmetry}
The clear asymmetry in brightness along the inner edge of the outer disk from the observations is naturally not produced with our symmetric disk model with spatially invariant dust properties. A slightly offset disk is one possibility for explaining the origin of the asymmetry and we explore this scenario in the following. We take our best axisymmetric model and slightly displace the star along the semi-major axis with respect to its original central position, while keeping the general disk structure unchanged. This is directly implemented into the radiative transfer code and not performed in a post-processing manner. A grid of additional models is computed, where the magnitude of the physical offset between the center of the T~Cha disk and the position of its host star is changed between 0.5 and 2.5\,au. We are only interested whether such a scenario is principally reliable, so we abstain from a fitting procedure. A value of $x = 2.1\,\mathrm{au}$, where $x$ is measured along the disk's semi-major axis, represents a reasonable match. The offset we apply is equivalent to a disk eccentricity of $e\approx0.07$. This way a brightness contrast between the east and west sides of 2 (Stokes $I$) and 3 (Stokes $Q_{\phi}$) can be reached (see Fig.~\ref{fig:rt_off}), consistent with the observational constraints. Other possible scenarios for the brightness asymmetry are discussed in Sect. \ref{subsec:discu_asym}.

\subsection{Point source analysis}
\label{subsec:ana_ps}

\begin{figure}
	\centering
	\centerline{
		\includegraphics[width=\columnwidth]{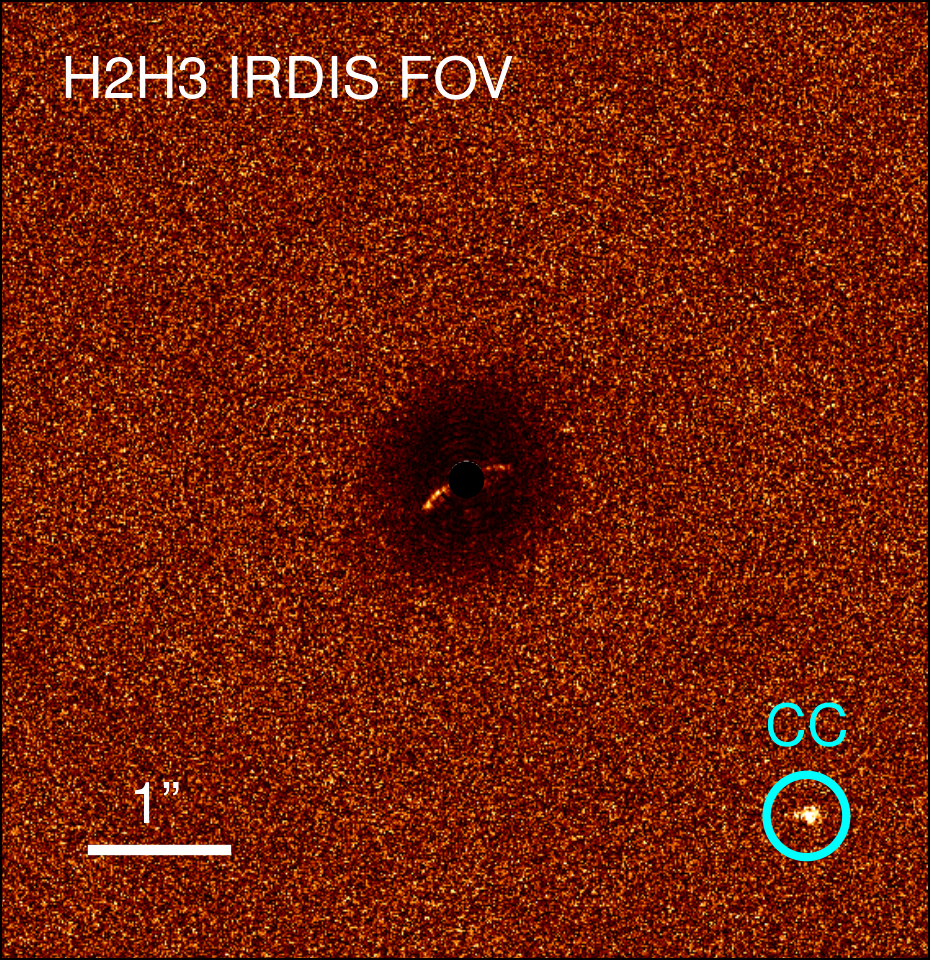}	
	}
	\caption{Signal-to-noise ratio map of the PCA reduction (\#2) of the IRDIS \textit{H2H3} data. The point source considered as a companion candidate (CC) is marked with a circle.}
	\label{fig:irdis_fov}	 
\end{figure}

\begin{figure}
	\centering
	\centerline{
		\includegraphics[width=\columnwidth]{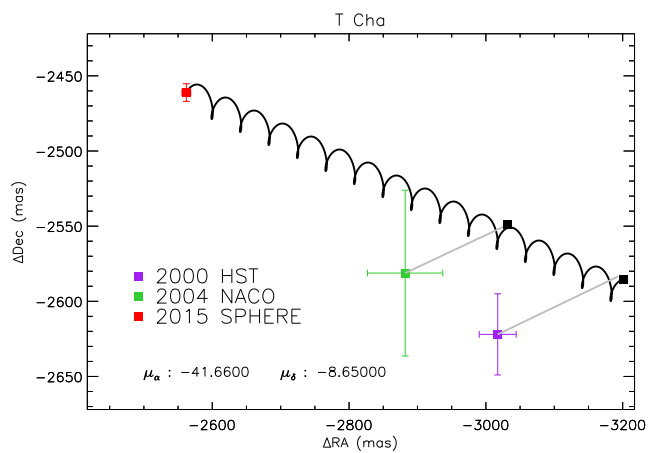}
	}
	\caption{Relative astrometry of the companion candidate labeled as `CC'
in Fig.~\ref{fig:irdis_fov} measured in SPHERE, NACO and HST data. The black solid line displays the motion of the companion if co-moving and the black squares are the positions expected at the time of HST and NACO observations.}
	\label{fig:pm}
\end{figure}

\begin{figure}
	\centering
	\centerline{
		\includegraphics[width=\columnwidth]{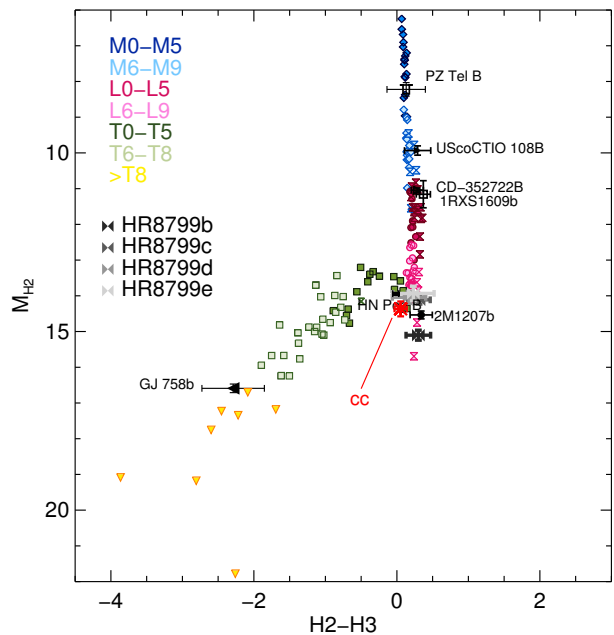}
	}
	\caption{Color-magnitude diagram displaying our candidate companion, which is marked in red and labeled with CC, compared to known substellar field (colored symbols) and young objects. Note that this plot assumes that CC is at the same distance as the star. Since CC is eventually classified as a background object based on a common proper motion test (cf. Fig.~\ref{fig:pm}), it is likely located much further.}
	\label{fig:shine_cmd}
\end{figure}

\begin{table}
	\caption{Astrometry and photometry relative to the star of the companion candidate in the T~Cha system}
	\label{tab:cc}
	\centering	
	\begin{tabular}{lcc}
		\hline\hline  
		\multicolumn{3}{c}{IRDIS companion candidate}\\
		\hline
	 	Filter & \textit{H2} & \textit{H3}\\
	 	$\lambda$ [$\mu$m] & 1.593 & 1.667\\
	 	Contrast [mag] & 11.65 $\pm$ 0.04 & 11.60 $\pm$ 0.04\\
	 	SNR & 31.1  & 31.4\\
	 	RA [mas] & -2560.3 $\pm$ 6.1 & -2562.8 $\pm$ 6.0\\ 
	 	DEC [mas] & -2460.1 $\pm$ 5.7 & -2461.29$\pm$ 5.7\\
	 	Separation [mas] & 3551.9 $\pm$ 8.2 & 3552.1 $\pm$ 8.1\\
	 	PA [deg] & 226.14 $\pm$ 0.18 & 226.16 $\pm$ 0.18\\
	\hline
	\end{tabular}
\end{table}

\begin{table}
  \caption{Relative astrometry of the companion candidate for different instrumental data}
  \label{tab:addlabel}  
  \centering
  \begin{tabular}{lcc}
  	\hline\hline
  	& NACO & SPHERE\\
  	\hline
    Date & 5 March 2004 & 30 May 2015\\
    JD & 2453070 & 2457173\\
    Separation [mas] & 3868.9$\pm$55.0 & 3552.6$\pm$10.7\\
    PA [deg] & 228.2$\pm$0.8 & 226.15$\pm$0.18\\
    \hline
    \end{tabular}
	\tablefoot{The NACO data was published in \cite{chauvin2010}.}
\end{table}

One candidate companion (CC) is detected in the IRDIS field of view (Fig.~\ref{fig:irdis_fov}), whereas no point-like sources are found in the IFS image. The speckle pattern is reduced in each frame of the sequence by subtracting an optimized reference image calculated by the TLOCI algorithm (\citealt{marois2010b}) implemented in SpeCal. We estimate the astrometry and photometry of this companion candidate using the calibrated unsaturated PSF (\citealt{galicher2011}) to remove biases. First, we roughly estimate the flux and position of the source in the TLOCI image. The SpeCal pipeline then creates a data cube of frames that only contain the unsaturated PSF at the candidate position on the detector, accounting for the field-of-view rotation in each frame. The TLOCI coefficients used to generate the TLOCI image where the candidate is detected are applied on the candidate data cube. The resulting frames are rotated to align north up. The median of these frames provides the estimation of the candidate image in the TLOCI image. We then adjust the estimated image subpixel position and its flux to minimize the integrated flux of the difference between the real and estimated candidate images. We use a 3$\times$FWHM diameter disk for the minimization. The 1$\sigma$ error bars are the required excursions in position or in flux to increase the minimum residual flux by a factor of $\sqrt{1.15}$ (cf. \citealt{galicher2016}, Galicher et al., in prep.). We empirically determine this factor running tests on sequences, in which we inject known fake planets. Using the calibrated unsaturated PSF, we also estimate the TLOCI throughput in all TLOCI sections following a procedure similar to the one used for the candidate position and flux estimation. The images were thus flux calibrated. The systematic errors for the astrometry of the detected companion candidate include the uncertainties on the pixel scale, North angle, frame centering using the satellite spots, accuracy of the IRDIS dithering procedure, anamorphic correction and SPHERE pupil offset angle in pupil-tracking mode (\citealt{vigan2016a,maire2016}). The calibration uses pixel scales of ($12.255\pm0.009$)\,mas/pix and ($12.251\pm0.009$)\,mas/pix for the \textit{H2} and \textit{H3} filters, respectively, and a true North offset of ($-1.712 \pm 0.063$)$^{\circ}$ is considered (\citealt{maire2016}).\\

CC is located at a separation of $(3.55\pm0.01)$\arcsec\ with contrast ($\Delta m_H=11.63\pm0.04$)\,mag (see Table \ref{tab:cc}). This same companion was already detected by \cite{chauvin2010} with $m_K=(11.4\pm0.1)$\,mag and is also present in HST data taken in coronagraphic mode with STIS in March 2000. Combining the new position measured from SPHERE with the old data we rule out this object as being gravitationally bound to T~Cha, because the motion observed over these 15 years is too large to be explained by a Keplerian orbit around this star; it is therefore a contaminant object. For completeness, given that the T~Cha proper motion is $\mu_\alpha=(-41.66\pm0.2)$\,mas/yr and $\mu_\delta=(-8.65\pm0.19)$\,mas/yr \citep{gaia2016a}, we also notice that CC has a high relative proper motion with respect to a background object (Fig.~\ref{fig:pm}). For completeness, we show the CMD in Fig.~\ref{fig:shine_cmd}. We note that this plot assumes that CC is at the same distance as T~Cha, since its actual distance is unknown. This is rather unlikely based on our previous conclusion that it is not physically associated with T~Cha. CC is likely located much further, and thus, likely intrinsically much brighter than an object at the L-T transition. Given the H2-H3 color~$\sim$0, we conclude that this object could be either a floating brown dwarf or a low mass star of the galactic thick disk or halo.\\

\subsection{Detection limits on substellar companion candidates}
\label{subsec:ana_limits}

The IRDIS detection limits for point sources are determined using the TLOCI data reduction. We estimate the 5$\sigma$ noise level, where $\sigma$ is the azimuthal robust deviation of the residual flux in annuli of $\lambda$/D
width rejecting pixels with no flux. Finally, the 5$\sigma$ noise levels are divided by the stellar flux estimated from the unsaturated images. The maximum contrast reached with IFS is obtained by applying the PCA technique. The contrast limits are estimated by an azimuthal standard deviation, that is, between pixels at the same separation from the star, for each angular separation, corrected by the star flux (obtained from the off-axis PSF images taken immediately before and after the coronagraphic observations) and the algorithm throughput (using synthetic companions injected into the data before the data processing as described above).\\

\begin{figure}
	\centering
	\centerline{
		\includegraphics[width=\columnwidth]{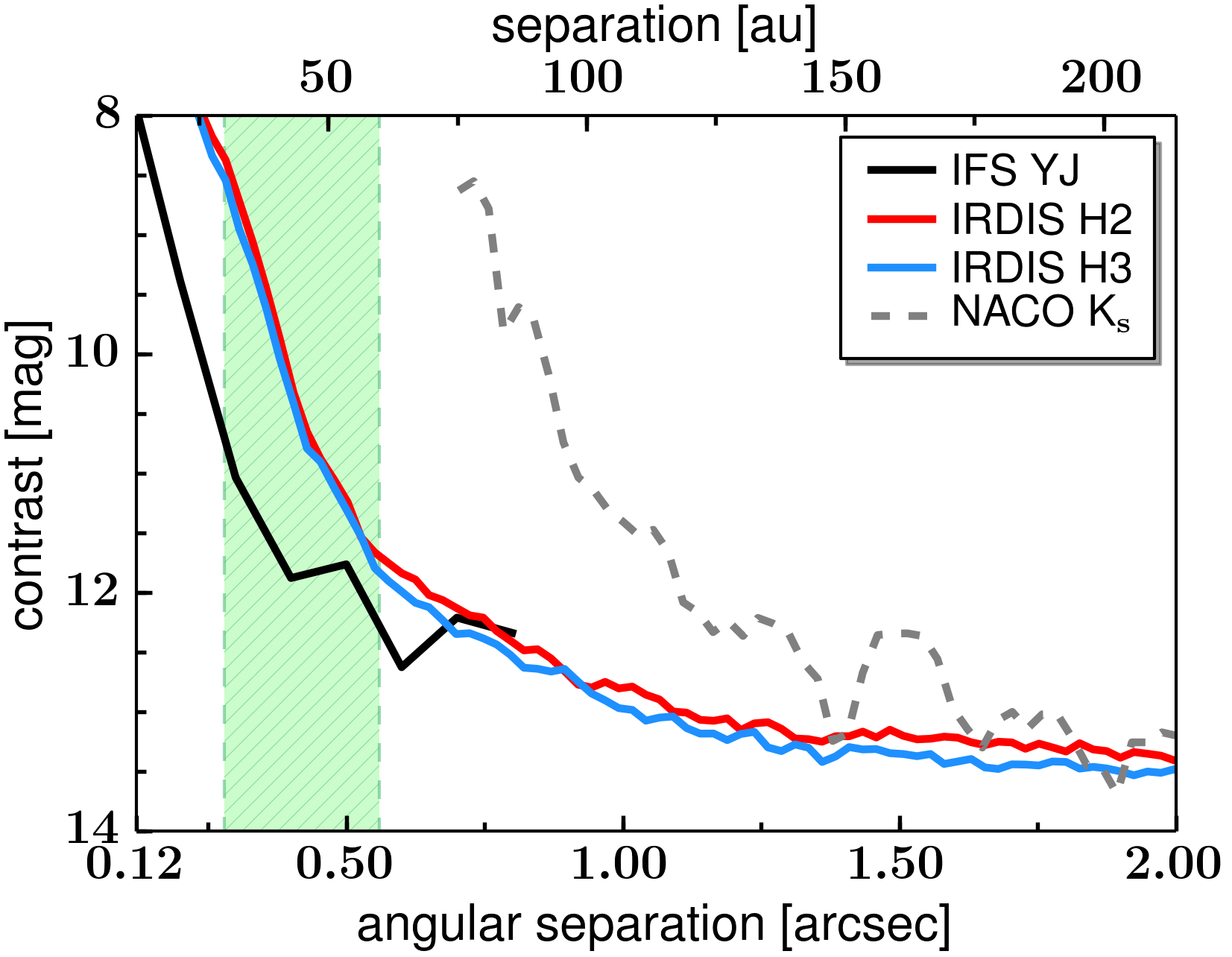}
	}
	\vspace*{0.3cm}
	\centerline{
		\includegraphics[width=\columnwidth]{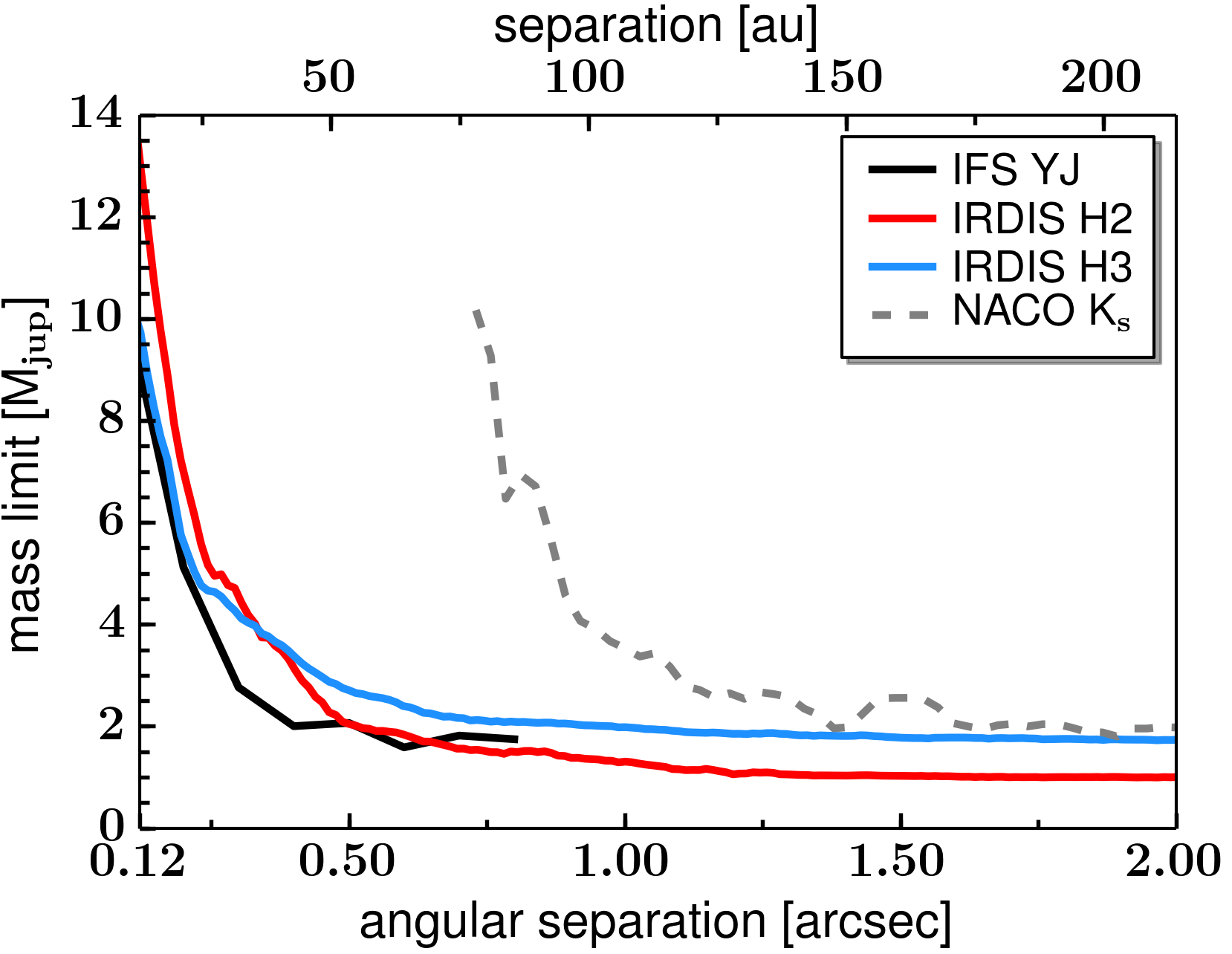}
	}
	\caption{Contrast curves and companion mass limits derived for IFS (black) after applying PCA, and for IRDIS \textit{H2} and \textit{H3} bands (red and blue, respectively) from the TLOCI reduction (\#3). Both curves have a lower cut at 0\farcs12. The detection limits from NACO \textit{K$_{\mathrm{s}}$} band data are given for comparison (gray dashed line, inner cut at 0\farcs7). The green striped rectangle denotes the area for which the contrast values might be slightly affected by the presence of the disk in scattered light.}
	\label{fig:shine_contrast}
\end{figure}

In Fig.~\ref{fig:shine_contrast} the contrast curves obtained for the different data sets are shown. The IRDIS data give a 5$\sigma$ contrast for a separation larger than 1.0\arcsec of greater than 12.5\,mag and 12.6\,mag in the \textit{H2} and \textit{H3} bands, respectively. Compared with NACO \textit{K$_{\mathrm{s}}$} band results (cf. \citealt{chauvin2010}), these observations are deeper by more than three magnitudes at a separation of 0\farcs7, that is, at the outer edge of the NACO coronagraph, while the contrast values at wider separations are comparable. IFS is deeper in contrast for separations closer than 0\farcs6 and gives 12\,mag in the \textit{YJ} band at a separation of $\sim$0\farcs7, assuming a gray contrast between the two objects.\\

Using the theoretical atmospheric models AMES-COND \citep{allard2003} we convert the contrast limits into upper limits on the mass of possible objects orbiting around T~Cha. These models are valid for T$_{\mathrm{eff}} < 1400\,$K and consider that the dust immediately rains out from the photosphere after its formation. We assume a system age of 7\,Myr (\citealt{torres2008}). This leads to a mass limit of $\sim$8.5\,$M_{\mathrm{jup}}$ in the innermost regions ($\sim$0\farcs1--0\farcs2), decreasing to $\sim$2\,$M_{\mathrm{jup}}$ for a separation between 0\farcs4 and 5.\arcsec. Our new SPHERE observations, therefore, improve the NACO mass limits especially up to $\sim$1.5\arcsec. Both, the contrast and mass curve are cut at 0\farcs12. The whole coronagraph system (apodizer, mask, stop) produces a radial transmission profile, which has not been accounted for in the derivation of the detection limits. The effect is visible at the region near the edge of the mask plus $\lambda/D$, thus, we exclude the inner 0\farcs12. Furthermore, we note that the detection limits represent an average value around the star, which might be affected by the disk signal at the location of the disk. However, we expect this effect to be small given the rather compact nature of the disk around T~Cha.

\section{Discussion}
\label{sec:discussion}

\subsection{Disk geometry}
\label{subsec:discu_geom}

Our analysis and modeling of the SPHERE data set confirms that the disk around T~Cha consists of an inner disk part and an outer disk part, separated by a cavity. Compared to the previous study by \citet{olofsson2013} we find the small dust cavity size to be larger by a factor of $\sim$2. Besides, it is even larger than the mm dust cavity of 20\,au estimated in \citet{huelamo2015}. This is rather unexpected, as the dust trapping scenario for transition disks is supposed to work such that bigger dust is trapped at a ring located outside of the small dust/gas cavity edge (see e.g., \citealt{pinilla2012,vandermarel2015}). This possible contradiction could be, however, due to uncertainties in the model fitting of data with low resolution by \citet{huelamo2015}. An inclination angle of $\sim$69$^{\circ}$ and PA of $\sim$114$^{\circ}$ best match our SPHERE observations, which is in agreement with \citet{huelamo2015}. Our new optical and near-infrared data do not, however, help us to constrain the outer disk radius. In our radiative transfer model we considered a tapered density profile for the dust density description of the outer disk, meaning that the surface density falls off gradually and hence, there is a smooth decrease of the dust mass per radius bin. However, simultaneously reproducing the gas and dust components of the disk remains challenging, and including this in our modeling effort is beyond the scope of this paper.

\subsection{Grain properties}
\label{subsec:discu_grains}

To simultaneously match the total intensity and polarimetric images obtained during our SPHERE observations, intermediate sized grains of $\sim$10\,$\mu$m must be present in the disk. This provides a better match with the observed properties of the disk than dust distributions covering several orders of magnitudes in size or a narrow distribution peaking at (sub-)micron size. This is in accordance with current grain growth models producing systematically larger grains, although we cannot guarantee that $\sim$10\,$\mu$m grains are located at the upper surface layer. By means of scattered light observations in the NIR we only trace the disk surface where the micron-sized grains are located for sure. Compact grains of a few tens of microns are expected to start settling down toward the disk midplane. The efficiency and timescale of vertical mixing depends, however, on the level of turbulence in disks which is still uncertain. With strong turbulence (high $\alpha$-viscosity, \citealt{shakura1973}) all grain sizes are better mixed. Thus, even larger grains can be present in the disk surface where they can contribute to the scattering. Furthermore, the amount of porosity of dust grains is unknown and still debated (e.g., \citealt{ossenkopf1993,dominik1997,kataoka2014}). For fractal aggregates with high porosity the phase function is supposed to differ (\citealt{tazaki2016}), which might alter our grain picture for T~Cha. A larger porosity for the same grain size might reduce the settling, where the size of the monomers still determines the absorption and scattering opacities. As shown by \cite{min2012}, the appearance of a disk in scattered light could be different depending on the fraction of fluffy aggregated dust particles compared to compact grains contained in the disk. We also note that very large grains (mm size) are indeed also expected to be present in the midplane in order to match the (sub-)mm data (cf. \citealt{huelamo2015}).

\subsection{Brightness asymmetry along the disk surface}
\label{subsec:discu_asym}

The intensity and polarized intensity distributions observed for T~Cha are asymmetric with respect to the minor axis of the disk. Similar brightness variation has also been detected in other disks, such as RY~Tau (\citealt{takami2013}) and AK~Sco (\citealt{janson2016}). In our radiative transfer modeling we explored the origin of the east-west asymmetry seen along the semi-major axis in the SPHERE observations by looking into the simplest possibility of a slightly offset disk. We approximate such an eccentric disk by calculating scattered light images of an azimuthally symmetric disk, but introducing an offset between the disk center and the star. A planetary companion on an eccentric orbit could shape the outer disk into an eccentric disk, causing the offset. Keeping T~Cha's stellar properties as the photon source in the radiative transfer code, but adding a positional offset, already reproduces well the observed asymmetry.\\

However, alternative explanations for the east-west brightness difference cannot be ruled out, and several effects may interact. Another idea is that an asymmetry in the inner disk or at the gap edges can lead to illumination effects helping to explain the dips in scattered light. The circumstellar disk around T~Cha may be actually still in an early stage of planetary formation. Thus, a dense dust clump formed in the inner, densest parts of the disk, or an already formed yet undetected planetary perturber below the detection limit, could cause this asymmetry. However, this scenario also raises the question of whether such an anisotropy is indeed stationary or moves with the local Keplerian velocity. A third scenario deals with spatially variant dust properties leading to a different scattering efficiency, which is especially related to grain size, structure, and composition. A possibility would be that unequal dust grain size distributions are present in the east and west wings of the disk, whose origin, however, remains unexplained. A fourth possible scenario leading to shadows in the outer disk is an inner disk significantly tilted with respect to the outer disk's plane. However, we find this scenario unlikely, since this arrangement would rather lead to relatively sharp, dark lanes, which are not apparent in our T~Cha images.

\section{Conclusions}
\label{sec:conclusions}

We have carried out VLT/SPHERE optical and NIR observations in polarimetric differential imaging mode with SPHERE/ZIMPOL in \textit{VBB} and SPHERE/IRDIS in \textit{H}-band of the evolved transition disk around the T~Tauri star T~Cha. Alongside the polarimetric observations, intensity images from IRDIS \textit{H2H3} dual-band imaging with simultaneous spectro-imaging with IFS in \textit{YJ}-band were obtained. The disk is clearly detected in all data sets presented in this work and resolved in scattered light with high angular resolution, allowing us to review the current understanding of the disk morphology and surface brightness. The basic structure of a classical transition disk previously reported by interferometric and (sub-)mm studies, has been confirmed. We developed a radiative transfer model of the disk including a truncated power-law surface density profile. The conclusions of this paper are summarized below.\\

\begin{enumerate}
	\item Our \textsc{RADMC-3D} radiative transfer model with updated disk parameters accounts well for the main geometry of the disk, the cavity, and the outer disk with its bright inner rim located at 0\farcs28 ($\sim$30\,au). This is significantly further out than previously estimated. A disk inclination of $\sim$69$^{\circ}$ and a position angle of $\sim$114$^{\circ}$ matches the SPHERE data sets best.\\
	
	\item We confirm that the dominant source of emission is forward scattered light from the near edge of the disk, given the high disk inclination. While small grains in the Rayleigh limit scatter photons rather isotropically and absorb very efficiently, large dust grains with sizes ($2\pi\mathrm{a}>\lambda$) have strong forward scattering properties. This demands a certain range of grain sizes to be present in the disk. We found that a power-law distribution with $a_{\mathrm{min}}=0.01\,\mu$m and $a_{\mathrm{max}}=1000\,\mu$m reproduces the total intensity observations well, but fails to be consistent with the polarimetric images. Thus, we propose a dominant grain size in the disk of $\sim$10$\,\mu$m. Such grains bring the desired amount of forward scattering and lead to a model that is in accordance with the complete SPHERE data set presented. We note that we restricted ourselves to the analysis of Mie theory and spherical compact grains. However, for aspherical aggregates with high porosity the phase function is supposed to differ, which might alter our grain picture for T~Cha.\\
	
	\item Our highly inclined disk model shows a significant $U_{\phi}$ signal at \textit{H}-band, which is in accordance with the observational $U_{\phi}/Q_{\phi}$ peak-to-peak value of 14\% and theoretical studies on multiple scattering events. The exact geometrical $U_{\phi}$ pattern observed with IRDIS is not reproduced, but the alternating structure of positive and negative lobes is well recognizable.\\
	
	\item The brightness asymmetry between the east and west sides can be reproduced with a slight offset of the star's position, representing a disk eccentricity of $e\approx0.07$. A planetary companion on an eccentric orbit could force the outer disk to become eccentric, causing this offset. However, a locally different grain size distribution and therefore a change of the scattering properties, or illumination effects due to asymmetric structures in the inner disk could also contribute to the brightness contrast observed.\\
	
	\item A previously known companion candidate is detected in the IRDIS field of view at a separation of $(3.54\pm0.01)$\arcsec\ with contrast $m_{\mathrm{H2}}=(11.63\pm0.04)$\,mag. We, however, rule out the possibility that this object is bound and, thus, conclude that it is not part of the T~Cha system.\\

	\item Our analysis rules out the presence of a companion with mass larger than $\sim$8.5\,$M_{\mathrm{jup}}$ between 0\farcs1 and 0\farcs3 from the central star, and larger than $\sim$2\,$M_{\mathrm{jup}}$ for wider separations. There could still be lower-mass planets in the outer disk regions and/or planets in the very inner disk.\\
	
\end{enumerate}


\begin{acknowledgements}
We would like to thank the ESO Paranal Staff for their support during the observations. We are very grateful to C.P.~Dullemond for insightful discussions. A.~P. is a member of the International Max Planck Research School for Astronomy and Cosmic Physics at Heidelberg University, IMPRS-HD, Germany. INAF-Osservatorio Astronomico di Padova acknowledges support from the "Progetti Premiali" funding scheme of the Italian Ministry of Education, University, and Research. M.~L., M.~B., F.~M. and C.~P. acknowledge funding from ANR of France under contract number ANR-16-CE31-0013. J.~O. acknowledges support from ALMA/Conicyt Project 31130027, and from the Millennium Nucleus RC130007 (Chilean Ministry of Economy). SPHERE is an instrument designed and built by a consortium consisting of IPAG (Grenoble, France), MPIA (Heidelberg, Germany), LAM (Marseille, France), LESIA (Paris, France), Laboratoire Lagrange (Nice, France), INAF-Osservatorio di Padova (Italy), Observatoire de Gen\`{e}ve (Switzerland), ETH Zurich (Switzerland), NOVA (Netherlands), ONERA (France) and ASTRON (Netherlands), in collaboration with ESO. SPHERE was funded by ESO, with additional contributions from CNRS (France), MPIA (Germany), INAF (Italy), FINES (Switzerland) and NOVA (Netherlands). SPHERE also received funding from the European Commission Sixth and Seventh Framework Programmes as part of the Optical Infrared Coordination Network for Astronomy (OPTICON) under grant number RII3-Ct-2004-001566 for FP6 (2004-2008), grant number 226604 for FP7 (2009-2012) and grant number 312430 for FP7 (2013-2016).
\end{acknowledgements}

%
%

\bibliographystyle{aa}
\bibliography{tcha}

\begin{appendix}

\section{Synthetic $Q_{\phi}$ image of model \#1}
\label{app:model1}

\begin{figure*}
	\centering
	\centerline{
		\includegraphics[height=0.33\textwidth]{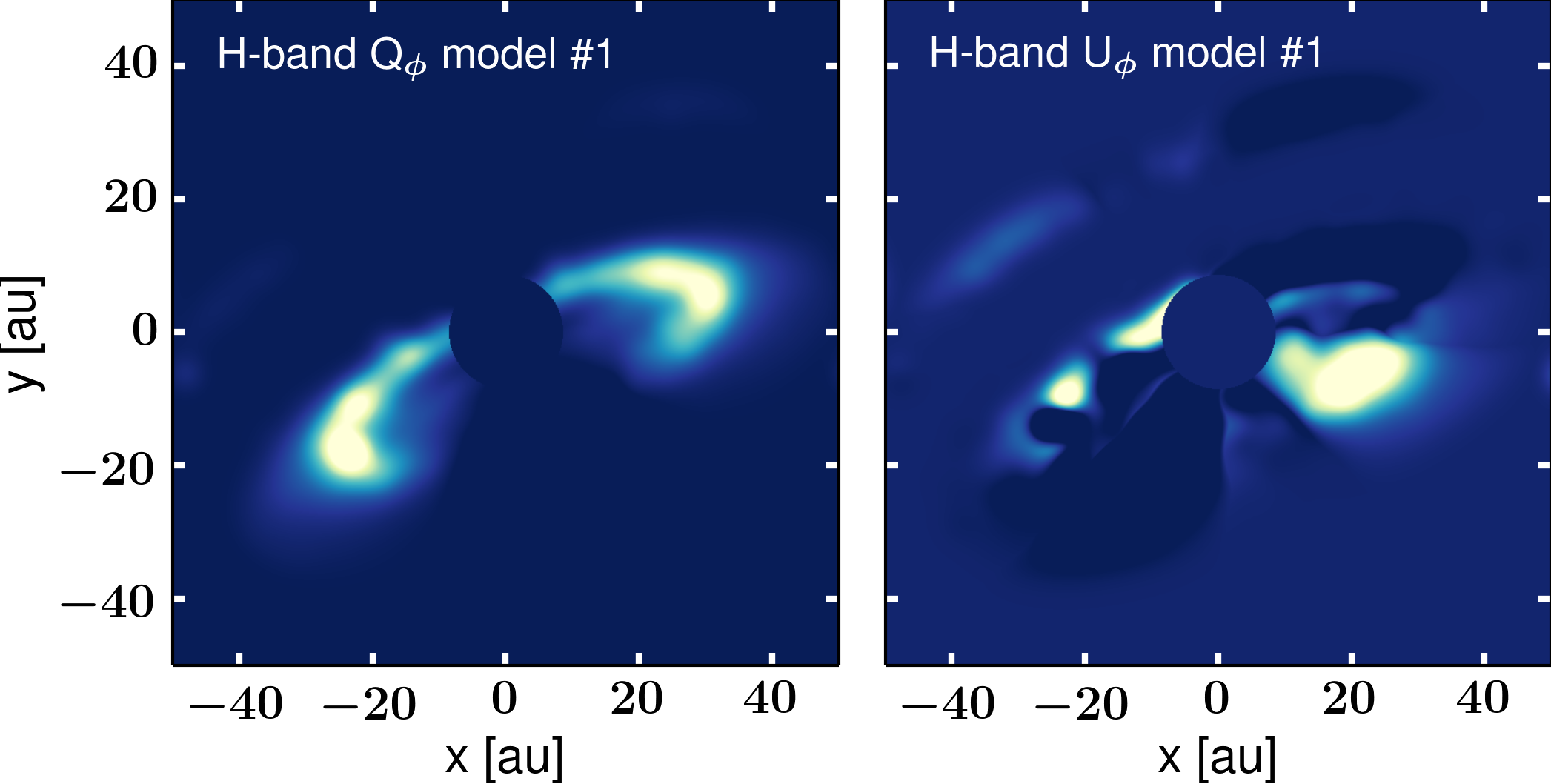}	
		\includegraphics[height=0.33\textwidth]{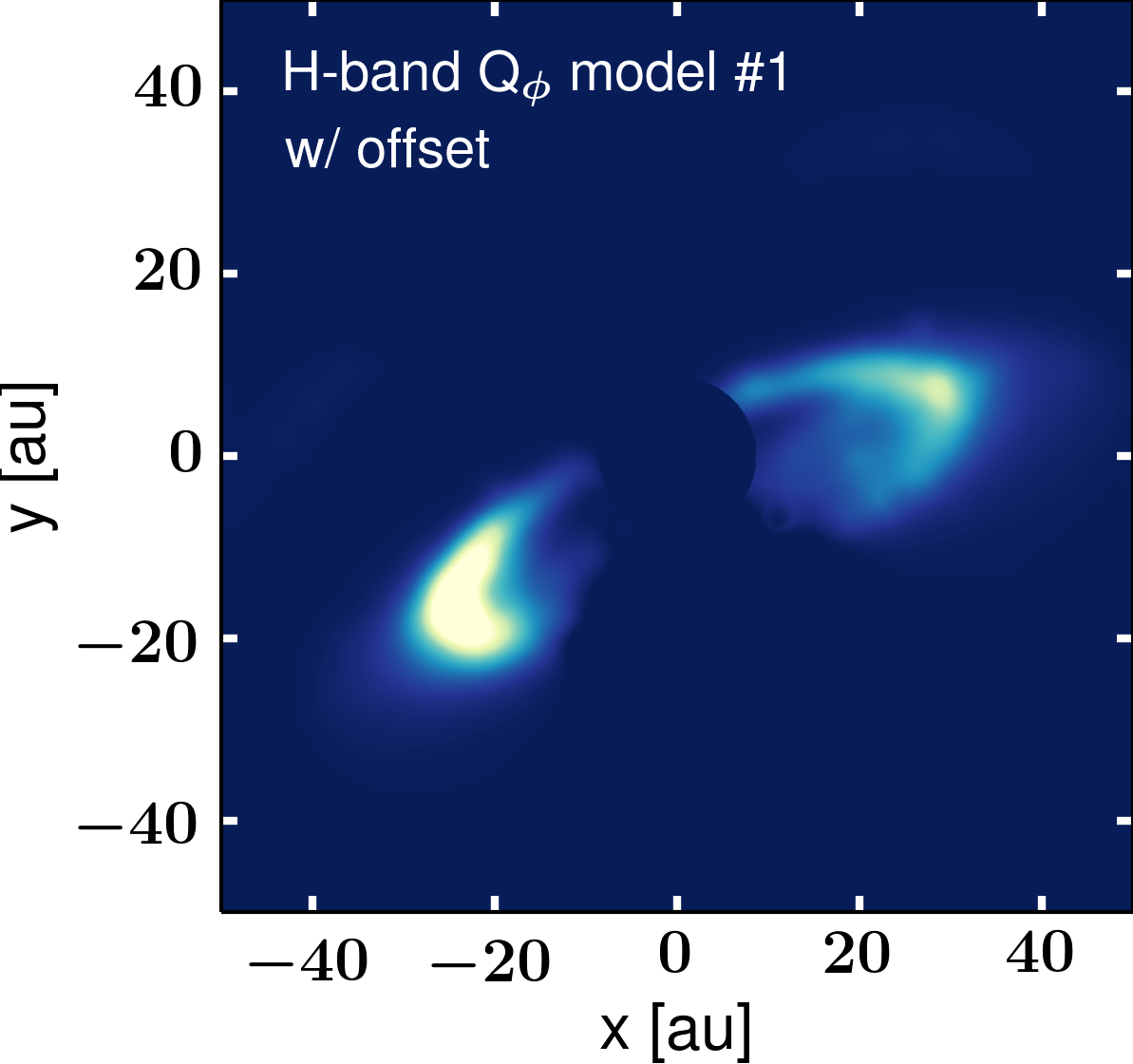}	
	}
	\caption{$Q_{\phi}$ (left) and $U_{\phi}$ (middle) images of model \#1 at \textit{H}-band. The right panel considers the same model, but the star is slightly offset from its original central position along the semi-major axis. The images are convolved with a Gaussian PSF with FWHM of 0\farcs04 (at 107\,pc distance). The color scale considers the same dynamical range as for the previous synthetic images.}
	\label{fig:mod1}	 
\end{figure*}

\end{appendix}

\end{document}